\documentclass[%
 reprint,
 amsmath,amssymb,
 aps,superscriptaddress
]{revtex4-2}

\usepackage{graphicx}
\usepackage{dcolumn}
\usepackage{bm}
\usepackage{tikz,xcolor}
\usepackage{physics}
\usepackage{float}
\usepackage{bibunits}
\usepackage{multirow}
\usepackage{hyperref}

\makeatletter
\def\maketitle{
\@author@finish
\title@column\titleblock@produce
\suppressfloats[t]}
\makeatother

\begin{document}
\begin{bibunit}[apsrev4-2]

\title{Carrier mobilities and electron-phonon interactions beyond DFT}

\author{Aleksandr Poliukhin}
\affiliation{Theory and Simulation of Materials (THEOS), École polytechnique fédérale de Lausanne, 1015 Lausanne, Switzerland}
\email{aleksandr.poliukhin@epfl.ch}
\author{Nicola Colonna}%
\affiliation{PSI Center for Scientific Computing, Theory and Data, 5232 Villigen PSI, Switzerland}%
\author{Francesco Libbi}
\affiliation{John A. Paulson School of Engineering and Applied Sciences, Harvard University, Cambridge, MA 02138, USA}%
\author{Samuel Poncé}
\affiliation{European Theoretical Spectroscopy Facility, Institute of Condensed Matter and Nanosciences, Université catholique de Louvain, Chemin des Étoiles 8, B-1348 Louvain-la-Neuve, Belgium}%
\affiliation{WEL Research Institute, avenue Pasteur 6, 1300 Wavre, Belgium}
\author{Nicola Marzari}
\affiliation{Theory and Simulation of Materials (THEOS), École polytechnique fédérale de Lausanne, 1015 Lausanne, Switzerland}
\affiliation{PSI Center for Scientific Computing, Theory and Data, 5232 Villigen PSI, Switzerland}%
\date{\today}
\begin{abstract}
Electron-phonon coupling is a key interaction that governs diverse physical processes such as carrier transport, superconductivity, and optical absorption. 
Calculating such interactions from first-principles with methods beyond density-functional theory  remains a challenge. 
We introduce here a finite-difference framework for computing electron-phonon couplings for any electronic structure method that provides eigenvalues and eigenvectors, and showcase applications for hybrid and Koopmans functionals, and $GW$ many-body perturbation theory. 
Our approach introduces a novel projectability scheme based on eigenvalue differences and bypasses many of the limitations of the direct finite difference methods. It also leverages symmetries to reduce the number of independent atomic displacements, thereby keeping computational costs manageable. 
This approach enables seamless integration with established first-principles codes for generating displaced supercells, performing Wannier interpolations, and evaluating transport properties. 
Applications to silicon and gallium arsenide show that advanced electronic-structure functionals predict different electron-phonon couplings and modify band curvatures, resulting in much more accurate estimates of intrinsic carrier drift mobilities and effective masses. 
In general, our method provides a robust and accessible framework for exploring electron-phonon interactions in complex materials with state-of-the-art electronic structure methods.
\end{abstract}
\maketitle

Electron-phonon interactions play a central role in determining fundamental materials' properties, such as electron and hole mobilities~\cite{li_ultra-fast_2019,ponce_first-principles_2020,claes_phonon-limited_2025}, superconductivity~\cite{bercx_charting_2025}, band renormalization~\cite{antonius_many-body_2014} and non-adiabatic effects~\cite{lazzeri_nonadiabatic_2006}. 
Accurate modeling of these interactions is essential for advancing technologies ranging from efficient electronic devices to novel superconducting materials.
First-principles calculations of electron-phonon couplings based on density-functional theory (DFT) have reached a maturity level that even allows high-throughput calculations~\cite{antonius_many-body_2014}.
A key step in these approaches is the evaluation of electron-phonon matrix elements, which quantify the effective interactions between electrons and phonons~\cite{Giustino_2017,bercx_charting_2025}.
Computationally, two approaches exist for this task: density-functional perturbation theory (DFPT)~\cite{Baroni_2001, gonze_dynamical_1997} and the finite difference (FD) method~\cite{Chang_1985, kresse_ab_1995, Vandenberghe_2014}. 
Recent work suggests machine learning as another promising alternative, as it allows accurate results when trained on relatively small datasets~\cite{zhong_accelerating_2024, li_deep-learning_2024}.
Although DFPT is generally a more computationally efficient approach than FD methods, as it offers favorable scaling and access to arbitrary phonon wavevectors $\mathbf{q}$, it requires dedicated and involved implementations.
For this reason, perturbative approaches based on DFPT have only recently been applied to specific beyond-DFT methods, such as DFT+U~\cite{hubbard_1991, hubbard_ep_2021, yang_first-principles_2025} and $G_0W_0$~\cite{gw_1985, li_electron-phonon_2019}. 
Up to now, most of the calculations on the electron-phonon-related properties have been performed using DFPT with semilocal DFT functionals and this method is considered to be the state of the art. 
%
Notwithstanding its success, discrepancies persist between first-principles predictions based on DFPT and experimental observations for many electron-phonon related properties~\cite{antonius_many-body_2014, tran_superconductor_2024, yuzbashyan_superconductivity_2022}.
This is partially due to the fact that Kohn-Sham (KS) DFT does not provide reliable quasiparticle energies and other excited-state properties~\cite{perdew_physical_1983,godby_trends_1987,marzari_electronic_2021}.
To address this, more advanced methods have been developed to improve the description of quasiparticle energies. 
These include the incorporation of a fraction of exact exchange~\cite{becke_new_1993,Heyd2003}, the enforcement of piecewise linearity conditions~\cite{dabo_koopmans_2010,colonna_koopmans-compliant_2019,borghi_koopmans-compliant_2014, nguyen_koopmans-compliant_2018,wing_band_2021,li_losc_2017}, and the inclusion of many-body effects~\cite{gw_1985, onida_electronic_2002,Golze2019}.
Improved electronic structure approaches were reported to affect electron-phonon properties~\cite{li_electron-phonon_2019,li_electron-phonon_2024}, but their impact on carrier mobility needs to be better understood.
For these reasons, there is a growing need for a more straightforward and general method to calculate electron-phonon couplings. 
For a long time, finite differences have been successfully used to predict phonon properties with any functional, and today, well-established codes exists that implement these developments~\cite{phonopy_2015, hjorth_larsen_atomic_2017,lin_first_2025}. 
However, the absence of community codes to perform a similar task for electron-phonon properties underscores the need for a more accessible and general approach. 
One of the primary objectives of this work is to establish an approach based on FD to calculate electron-phonon matrix elements starting from any arbitrary electronic-structure method that provides eigenvectors and eigenvalues of the Hamiltonian for displaced systems. 
This approach and the code that implements it aim to fill the gap in the current software landscape. 
We illustrate our approach using \textsc{Quantum ESPRESSO}~\cite{Giannozzi_2009, Giannozzi_2017}, \textsc{Koopmans}~\cite{linscott_koopmans_2023}, and \textsc{Yambo}~\cite{marini_yambo_2009} as {\textit ab-initio} electronic structure engines.
Furthermore, we provide an interface to the \textsc{EPW} software~\cite{ponce_epw_2016, lee_electronphonon_2023} and compute the transport properties of Si and GaAs to assess the effect that using hybrid and Koopmans functionals, and $GW$ many-body perturbation theory  have on transport properties.

\section{Results}
The central quantity that describes electron-phonon interactions arises from the expansion of the effective potential that acts on the electrons due to the motion of atoms, which is given by:
\begin{equation}
    \label{eq:disp_pot}
    \hat{H}^{\rm e-ph} = \sum_{\kappa\alpha l} \frac{\partial \hat{V}}{\partial \tau_{\kappa\alpha l}} \delta \tau_{\kappa\alpha l},
\end{equation}
where $\hat{V}$ is the effective potential, $\tau_{\kappa\alpha l} = \mathbf{R}_l +  \tau_{\kappa\alpha}$ is the $\alpha$ Cartesian component of the position of the atom $\kappa$ in a unit cell located at position $\mathbf{R}_l$. 
The matrix element of this operator describes the probability that an electron transitions from an initial to a final state because of the interaction with the atomic displacement $\delta \tau_{\kappa\alpha l}$. 
Expanding the atomic displacement $\delta \tau_{\kappa\alpha l}$ on the phonon basis, it is possible to define the electron-phonon matrix element $g_{mn\nu}(\textbf{k},\textbf{q})$ as:
\begin{multline}\label{eq:g_full}
    g_{mn\nu}(\textbf{k},\textbf{q}) \equiv \\
     \sum_{\kappa\alpha l} \sqrt{\dfrac{\hbar}{2 M_\kappa \omega_{\textbf{q}\nu}}} e^{i\textbf{q} \cdot \textbf{R}_l} e_{\kappa\alpha\textbf{q}\nu}  \bra{\psi_{m\textbf{k}+\textbf{q}}} \dfrac{\partial V}{\partial \tau_{\kappa\alpha l}} \ket{{\psi_{n\textbf{k}}}},
\end{multline}
 where $\omega_{\textbf{q}\nu}$, $e_{\kappa\alpha\textbf{q}\nu}$ are the phonon frequencies and eigenvectors respectively, $\psi_{n\textbf{k}}$ is the one-particle wave function, and $\partial V / \partial \tau_{\kappa\alpha l}$ is the change of potential with respect to the displacement. 

Standard FD approaches require one to evaluate the change in effective potential due to atomic displacement by using finite differences in a displaced supercell commensurate with the phonon \textbf{q}-vector of interest.
However, in pseudopotential codes, there is a non-local part of the potential that is treated analytically and thus would require an additional implementation when different types of pseudopotential are used, such as USPP~\cite{vanderbilt_ultrasof_1990} or PAW~\cite{blochl_paw_1994}.
For norm-conserving pseudpotentials the expression is given in Sec.~S1,  Eq.~\eqref{eq:elph:NL} of the SI~\cite{supplement}.
Moreover, in beyond-DFT approaches, such as hybrids, Koopmans, or $G_0W_0$, the effective potential is also a more complex operator that requires additional treatment, such as saving the unique potential for every orbital, which introduces additional memory constraints on the calculations.
These challenges prevent the FD method from serving as a general-purpose approach for arbitrary electronic-structure functionals, as it would require distinct treatments for DFT and beyond-DFT methods, as well as for different pseudopotential schemes.
To enable a unified approach to the calculation of electron-phonon couplings across different functionals, we propose here an alternative and general method that provides an easy and elegant solution to the challenge discussed above.

\subsection{Projectability approach based on eigenvalues}
The derivative in Eq.~\eqref{eq:g_full} can be seen as the derivative of the Hamiltonian.
Since the kinetic energy does not depend on the atomic position, the braket can be written as: 
\begin{align}\label{eq:elph_project_sc}
    \bra{\psi_{m\textbf{k}+\textbf{q}}}  \dfrac{\partial \hat{V}}{\partial \tau_{\kappa\alpha l}}  \ket{\psi_{n\textbf{k}}} =& \bra{\psi_{m\textbf{k}+\textbf{q}}} \dfrac{\partial \hat{H}}{\partial \tau_{\kappa\alpha l}} \ket{\psi_{n\textbf{k}}} \nonumber \\ 
    \approx &    \bra{\psi_{m\textbf{k}+\textbf{q}}} \dfrac{\hat{H}(\tau_{\kappa\alpha l})}{2\tau_{\kappa\alpha l}} \ket{\psi_{n\textbf{k}}} \nonumber\\
    & - \bra{\psi_{m\textbf{k}+\textbf{q}}} \dfrac{\hat{H}(-\tau_{\kappa\alpha l})}{2\tau_{\kappa\alpha l}} \ket{\psi_{n\textbf{k}}}.  
\end{align}
Using the first quantization representation of the Hamiltonian $\hat{H} = \sum_j \varepsilon_j \ket{\psi_j} \bra{\psi_j}$, one arrives at
\begin{align}
    \label{eq:projectability}
    \bra{\psi_{m\textbf{k}+\textbf{q}}} \dfrac{\partial V}{\partial \tau_{\kappa\alpha l}} \ket{\psi_{n\textbf{k}}} =&  \frac{\tau^{-1}_{\kappa\alpha l}}{2}\Big(\sum_{j} \varepsilon^+_j u_{m\textbf{k+q}j}^{*+}  u^{+}_{jn \textbf{k}}  \nonumber \\
    &-  \sum_{j} \varepsilon^-_j u_{m\textbf{k+q}j }^{*-} u^{-}_{jn \textbf{k}}\Big), \\
    u^{+}_{jn\textbf{k}} =& \bra{\psi^+_j} \ket{\psi_{n\textbf{k}}} \label{eq:projectability_u} \\
    u^{-}_{jn\textbf{k}} =& \bra{\psi^-_j} \ket{\psi_{n\textbf{k}}},
\end{align}
where $\psi^{\pm}_j$ and $\varepsilon^{\pm}_j$ are the wavefunction and eigenvalues of the perturbed supercell with band index $j$ and $\mathbf{k}$=$\boldsymbol{\Gamma}$, and the $+$ and $-$ symbols represent positive and negative atomic displacement in Cartesian coordinates, respectively.  
Equation~\eqref{eq:projectability} represents the eigenvalue projectability approach for calculating electron-phonon matrix elements. 
Similar ideas were proposed for localized basis sets, such as LCAO~\cite{croy_dftbephy_2023} or Wannier functions~\cite{engel_electron-phonon_2020,wang_accurate_2024,ep_wannier_2022}, which have advantages in localizing perturbations in real space. 
However, constructing the perturbed Wannier functions in the supercells might not generally guarantee smoothness with respect to the unperturbed calculation. 
Moreover, the limiting factor in all the finite difference approaches is the self-consistent calculation of the ground state of the displaced supercell, meaning that any postprocessing step (as the one in Eq.~\eqref{eq:projectability}) represents just a fraction of the total computational time.   

The main advantage of Eq.~\eqref{eq:projectability} is that it only requires information about eigenvectors and eigenvalues of the pristine and displaced system, making the approach general and applicable to any beyond-DFT method as long as they deliver the quantities mentioned above.
Established approaches~\cite{antonius_many-body_2014} allow for estimating electron-phonon coupling for specific points in the Brillouin zone using eigenvalue differences. 
Here, instead, by performing supercell calculations, one can automatically get information about electron-phonon matrix elements at any $\mathbf{q}$-point commensurate with the size of the displaced supercell. 

Two approximations are used in the derivation of Eq.~\eqref{eq:projectability}: a finite difference formula for the evaluation of the change in the effective Hamiltonian, and the resolution of the identity that in practical implementations needs to be approximated with a finite set of eigenstates. 
The first approximation requires a convergence study with respect to the finite-difference step. 
Detailed information on the convergence of the FD approach can be found in Fig.~\ref{fig:dfpt_converge}, Fig.~\ref{fig:converge_1ep}, and Fig.~\ref{fig:epw_converge} of the SI~\cite{supplement}.
To control the quality of the second approximation, we introduce the following unitary conditions: 
\begin{align}\label{eq:projectability_condition}
    P^{+}_{nm\textbf{k}}(N^{\rm max}) =& \sum^{N^{\rm max}}_{j} \bra{\psi_{n\textbf{k}}}\ket{\psi^{+}_j} \bra{\psi^{+}_j}\ket{\psi_{m\textbf{k}}} \approx \delta_{mn}\\
    P^{+}_{nn\textbf{k}}(N^{\rm max}) =& \sum^{N^{\rm max}}_{j} |u^{+}_{jn\textbf{k}}|^2 \leq 1, \label{eq:projectability_condition_2}
\end{align}
\begin{figure}
    \includegraphics[width=0.96\linewidth]{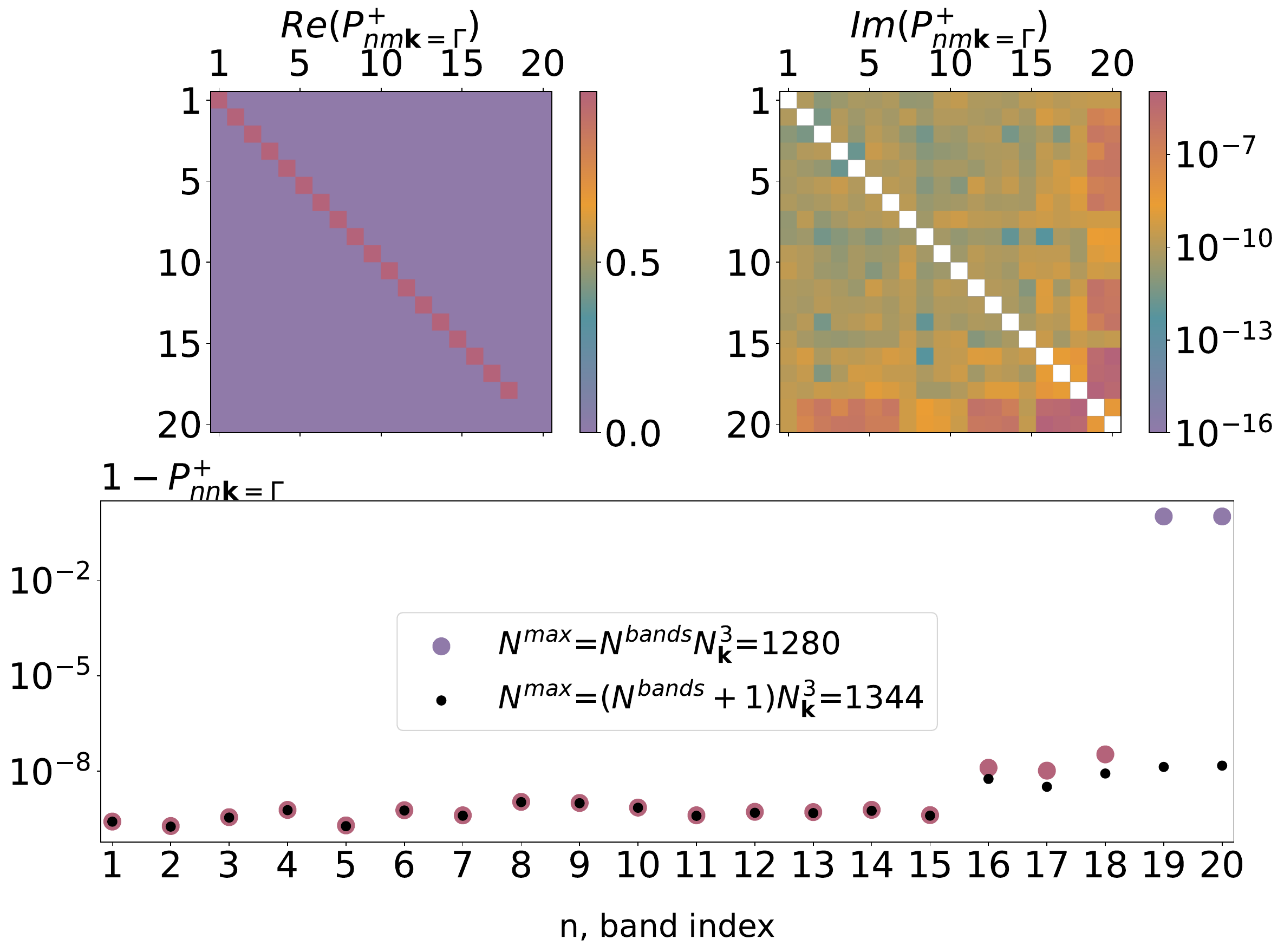}
    \caption{\label{fig:projectability_dft} Verification of projectability condition in Eq.~\eqref{eq:projectability_condition} and ~\eqref{eq:projectability_condition_2} for the 20 unperturbed states of silicon at \textbf{k}=$\Gamma$. 
    The corresponding displaced supercell has dimensions 4$\times$4$\times$4 and $N^{\rm max}$ is set to 1280. 
    The top left and top right are real and imaginary parts of the $P^{+}_{nm\textbf{k}}$ matrix that should be close to identity. 
    In this case, a good projectability holds only for the first 18 states. 
    On the bottom panel, the diagonal part of the matrix is represented for the case when $N^{\rm max}$ = 1280 and $N^{\rm max}$ = 1344 are considered. 
    Including all degenerate states ($n$ = 19, 20, 21) allows for the fulfillment of the condition.
    }
\end{figure}
\noindent where the sum runs over the eigenstates of the displaced system in the supercell setup up to a chosen threshold $N^{\rm max}$.
Using perturbation theory for small displacements, it can be shown (see Eqs.~\eqref{eq:project_err1} and \eqref{eq:project_err2} in SI~\cite{supplement}) that if the unperturbed state $n\textbf{k}$ has a corresponding perturbed state $j$ in Eq.~\eqref{eq:projectability_condition}, the error is $\mathcal{O}(\tau^2)$. 
Thus, if $N^{\rm bands}$ and  $N_{\textbf{k}}$ are the number of bands and the $\textbf{k}$-point sampling used for the pristine unit cell calculation repsectively, then one needs at least $N^{\rm max}=N^{\rm bands}N^3_{\textbf{k}}$ states in the displaced supercell calculation with $\boldsymbol{\Gamma}$ sampling.
However, in case the last unperturbed state is degenerate, more bands should be included in $N^{\rm max}$.
The numerical evidence of this is presented in Fig.~\ref{fig:projectability_dft} where Eq.~\eqref{eq:projectability_condition}, \eqref{eq:projectability_condition_2} are illustrated for the 4$\times$4$\times$4 supercell of Si. 
In the first case, when we consider $N^{\rm max}$ = 1280, which corresponds to 20 unperturbed states, the projectability condition fails for the last 2 unperturbed states. 
This is because the unperturbed states 19, 20, and 21 are degenerate and only two of the three states were included in the sum. 
Increasing $N^{\rm max}$ to 1344 (21 unperturbed states) brings the projectability condition in Eq.~\eqref{eq:projectability_condition} back to a value close to 1.
For an efficient evaluation of Eq.~\eqref{eq:projectability} one could also realize that most of the values in $u_{jm\textbf{k}}^{\pm}$ are negligible since they represent the overlaps between perturbed and unperturbed wavefunctions with different quantum numbers. 
Note that the brakets in Eq.~\eqref{eq:projectability_u} are calculated between the unperturbed wave functions initially constructed in the unit cell and the wave functions of the supercell. 
For a consistent evaluation of the integral, unperturbed wave functions need to be unfolded in the supercell. After the unfolding, a Fourier transformation to reciprocal space can be performed, allowing for an efficient calculation of the brakets. 
We note that this unfolding procedure also allows us to keep track of the dependence of the electron-phonon matrix elements on \textbf{k} and \textbf{q} vectors in the primitive cell.

\subsection{Symmetries of displaced calculations}
In order to evaluate electron-phonon matrix elements at finite \textbf{k} and \textbf{q} point, according to Eq.~\eqref{eq:g_full}, we need to displace each atom in 3 Cartesian directions, performing in total 3$N^{\rm at}N^{\rm uc}$ supercell calculations, where $N^{\rm at}$ is the number of atoms in the unit cell, and $N^{\rm uc}$ is number of unit cells in the supercell. 
Using the periodicity of the system, we have that Eq.~\eqref{eq:g_full} becomes~\cite{gunst_first-principles_2016}: 
\begin{multline}
    g_{mn\nu}(\textbf{k},\textbf{q}) = \\
     N^{uc}\sum_{\kappa\alpha} \sqrt{\dfrac{\hbar}{2 M_\kappa \omega_{\textbf{q}\nu}}}  e_{\kappa\alpha\textbf{q}\nu}  \bra{\psi_{m\textbf{k}+\textbf{q}}} \dfrac{\partial V}{\partial \tau_{\kappa\alpha 0}} \ket{{\psi_{n\textbf{k}}}},
\end{multline}
where importantly $\partial V / \partial \tau_{\kappa\alpha}$ needs to be evaluated only for a single reference unit cell in the supercell, reducing the number of displacements to 3$N^{\rm at}$. 
Evaluating the derivative with a second-order finite-difference schem requires displacing the atoms in the $+$ and $-$ directions and increases the number of supercell calculations to 6$N^{\rm at}$.
However, it is known~\cite{Togo_2023} that symmetry considerations for the case of phonon calculations allow reducing the number of displacements to $N^{\rm in}$, where $N^{\rm in}$ is the number of inequivalent atoms in the unit cell. 
For phonons, forces are calculated only for specific displacements and then rotated to obtain a complete set of displacements. In a perfect analogy, we consider a set of displacements $\{{\lambda}\}$ that transform under the following conditions:
\begin{align}\label{eq:symmetries}
        \hat{S}\lambda =& \lambda' \\
        \hat{S}R_\kappa =& R_{\kappa'},
\end{align}
where $\hat{S}$ represents the operator of space group symmetries that maps atom $\kappa$ to atom $\kappa'$. 
In this case, the potential evaluated at displacement $\lambda'$ is related to that for displacement $\lambda$ by a basis transformation:
\begin{equation}
    \label{eq:potential_rot}
     V^{R_{\kappa'}+\lambda'}(\mathbf{r}) =V^{R_{\kappa}+\lambda}(\hat{S}^{-1}\mathbf{r}).
\end{equation}
Eq.~\eqref{eq:potential_rot} also implies that the Hamiltonian satisfies the same condition because the kinetic energy is invariant under unitary transformations of the basis. 
This means that the effective equations for the displacements of $\lambda$ and $\lambda'$ produce the same eigenvalues and are connected by:
\begin{align}
   \label{eq:ks_displacement_1}
   \hat{H}^{R_{\kappa'}+\lambda'}(\mathbf{r}) \psi^{R_{\kappa'}+\lambda'}_{n}(\mathbf{r}) =& \varepsilon_{n} \psi^{R_{\kappa'}+\lambda'}_{n}(\mathbf{r})\\
   \hat{H}^{R_{\kappa}+\lambda}(\hat{S}^{-1}\mathbf{r}) \psi^{R_{\kappa}+\lambda}_{n}(\hat{S}^{-1}\mathbf{r}) =& \varepsilon_{n}  \psi^{R_{\kappa}+\lambda}_{n}( \hat{S}^{-1}\mathbf{r}).
\end{align}
Therefore, one performs only $N^{\rm in}$ displacements and then obtains the wavefunctions of the other displacements by symmetry $\psi^{R_{\kappa'}+\hat{S}\lambda}_{n}(\mathbf{r}) = \psi^{R_{\kappa}+\lambda}_{n}( \hat{S}^{-1}\mathbf{r})$. 
An important caveat is that the displacements obtained by applying the symmetry operations have to be linearly independent when the same atom $\kappa$ is considered.
Indeed, Eq.~\eqref{eq:g_full} is written as the sum of derivatives of the potential along Cartesian displacements that would be connected to the set of displacement $\lambda$, $\mu$, $\nu$ obtained by the symmetry operations:
\begin{gather}
    \label{eq:sym_rot}
     \begin{bmatrix} \lambda_x & \lambda_y & \lambda_z \\  \mu_x & \mu_y & \mu_z \\ \nu_x & \nu_y & \nu_z \end{bmatrix}
     \begin{bmatrix} \frac{\partial V}{\partial x_{\kappa}}\\  \frac{\partial V}{\partial y_{\kappa}} \\ \frac{\partial V}{\partial z_{\kappa}} \end{bmatrix}
     =
    \begin{bmatrix} \frac{\partial V}{\partial \lambda_\kappa}\\  \frac{\partial V}{\partial \mu_\kappa} \\ \frac{\partial V}{\partial \nu_\kappa} \end{bmatrix},
\end{gather}
where the matrix on the left-hand side represents the transformation matrix between 3 independent displacements and 3 Cartesian displacements of atom $\kappa$. 
If any displacements are linearly dependent, the equation becomes undetermined.

\subsection{Electron-phonon coupling with realistic electrons}
\begin{figure}
    \includegraphics[width=0.99\linewidth]{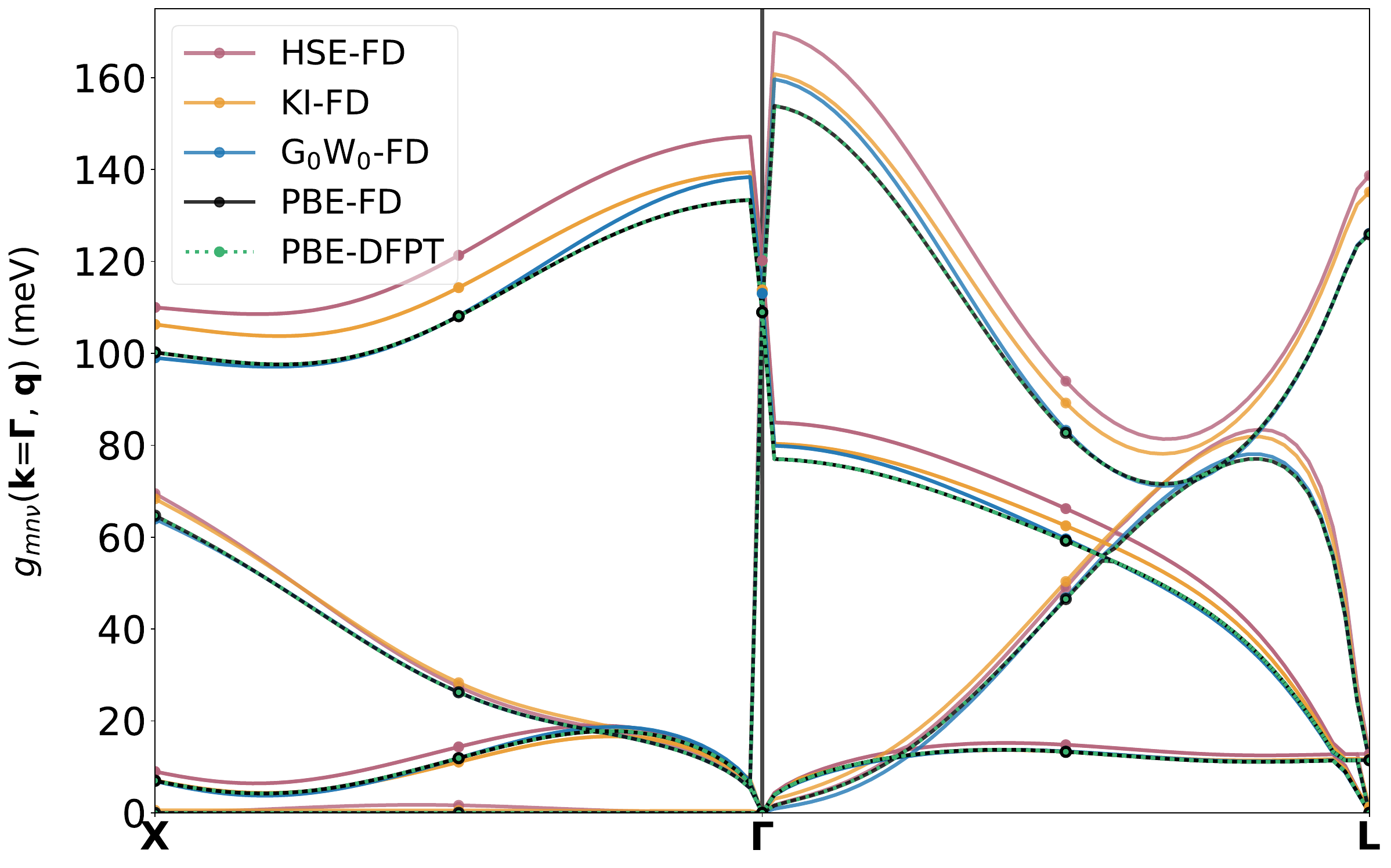}
    \caption{\label{fig:el_ph_fine}
 Interpolated electron-phonon matrix elements of Si along a $\textbf{q}$-momentum path at $\textbf{k}$ = $\boldsymbol{\Gamma}$ for the electronic states of the second valence band ($m$=$n$=2). 
    All the results are interpolated with \textsc{EPW} using a coarse 4$\times$4$\times$4 $\mathbf{k/q}$ supercell.
    We show the result from a density-functional perturbation theory (DFPT) approach, interpolated from the same coarse grid, using a dashed green line. 
    The direct results obtained using FD or DFPT on the 4$\times$4$\times$4 $\mathbf{k/q}$-point grid are highlighted with circles.
    The PBE supercell finite difference (FD) and DFPT approach match perfectly, validating the approach.
    }
\end{figure}
%
\begin{figure*}
    \includegraphics[width=1.0\linewidth]{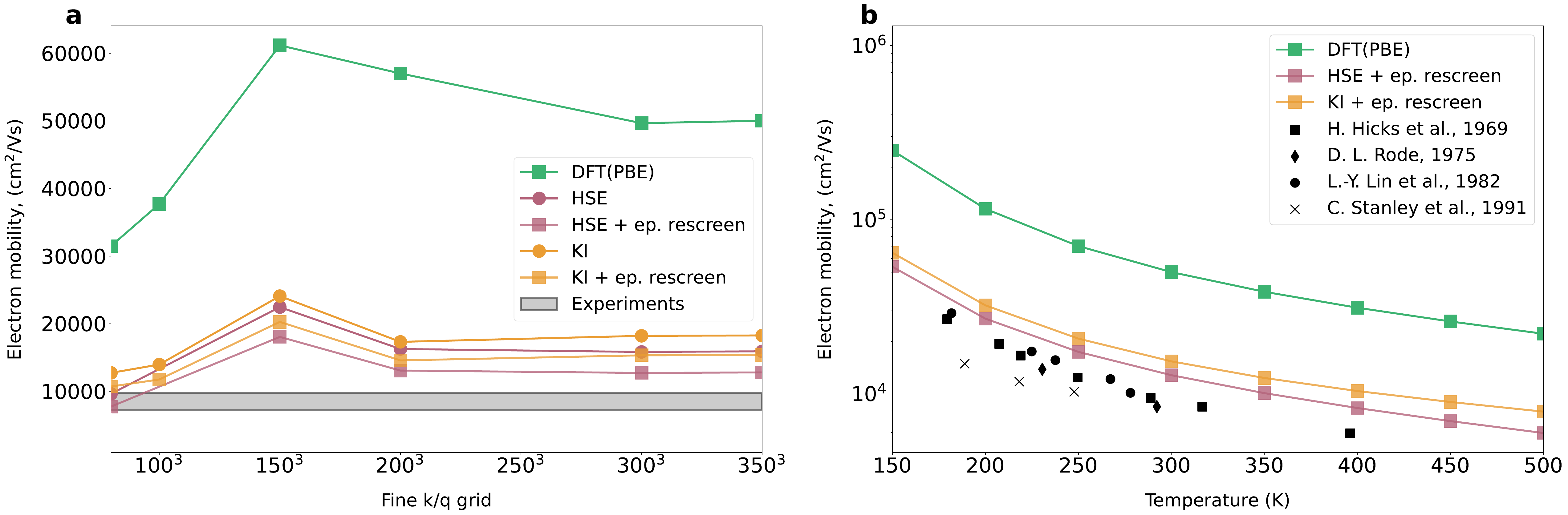}
    \caption{\label{fig:mob_gaas}  Mobility with beyond-DFT approaches. \textbf{a} Convergence of electron drift mobility of GaAs at room temperature with different functionals.
    The results with circles are obtained by only changing the band structure. In contrast, the results with squares also consider the effects of beyond-DFT functionals on electron-phonon coupling by introducing average rescreening of the electron-phonon matrix. 
    The experimental results~\cite{madelung2003, hicks1969, rode1971} are in good agreement with HSE and KI functionals.
    \textbf{b} Temperature dependence of electron drift mobility of GaAs with different functionals. 
    Black markers represent experimental results~\cite{ hicks_high_1969, rode_semicond_1975, lan-ying_vapour_1982, staneley_4_1991} that lie very close to beyond-DFT values.}
\end{figure*}
To show the simplicity and flexibility of our approach, 
we study the electron-phonon coupling of silicon at different levels of theory: DFT, HSE, Koopmans (KI), and $G_0W_0$. 
Since most observables require a dense \textbf{k} and \textbf{q} momentum integration of the electron-phonon matrix elements, we initially perform a calculation with a relatively small supercell and then use a Wannier interpolation approach~\cite{marzari_maximally_1997, Giustino_2007, marzari_maximally_2012} to obtain smooth functions. 
The interpolated electron-phonon coupling of silicon for the band indices $m$=$n$=2 along high symmetry line $\boldsymbol{X} -\boldsymbol{\Gamma}-\boldsymbol{L}$ is shown in Fig.~\ref{fig:el_ph_fine}. 

The perfect agreement between DFPT and FD results for the standard semilocal PBE functional validate the projectability approach proposed here. 
Beyond DFT approaches show an increase in electron-phonon coupling compared to semilocal DFT with HSE having an enhancement of 9\%. 
This increase could be explained by the fact that DFT underestimates the electronic bandgap and overestimates the screening, leading to an overestimated dielectric constant. 
Since the electron-phonon matrix elements are inversely proportional to the screening~\cite{Giustino_2017}, such underestimation is expected. 
It is also important to stress that different electronic-structure methods produce different phonon spectra, as shown for the case of silicon in Fig.~\ref{fig:ph_epw} of SI where the HSE functional is shown to increase the phonon frequencies on average by 5\%.
However, the KI functionals yields the same phonon dispersion as the DFT functional it is based on. 
This is because for insulating or semiconducting systems at integer electron number, the KI functional preserves the energetics (total energy and static derivatives of the total energy) of the underlying density-functional approximation.
Therefore, KI forces are identical to DFT forces with the same interatomic force constant.
Phonon properties with $G_0W_0$ are also kept at the DFT level treating the eigenvalues correction in a perturbative way \cite{li_electron-phonon_2019}. 
We note that in general DFT produces phonon dispersions in very good agreement with experiment~\cite{giannozzi_ab_1991}.

We find that the increase in electron-phonon matrix element is quite similar across phonon \textbf{q} and electron \textbf{k} (quasi-)momenta, see Fig.~\ref{fig:el_ph_coarse}.
The only exception in this case is $G_0W_0$ that tends to have a similar increase in coupling as in the Koopmans case close to $\Gamma$, but provides a negligible correction at finite \textbf{q}.
Overall, the application of $G_0W_0$ corrections to the electron-phonon coupling still provides smooth corrections both as a function of \textbf{q} and \textbf{k} (see Fig.~\ref{fig:el_ph_coarse} of the SI). 
The small values of the correction to the coupling obtained here are in good agreement with previous works on band renormalization in silicon~\cite{karsai_electronphonon_2018}, although other calculations in the literature have reported more significant renormalization effects~\cite{monserrat_correlation_2016}.
In the case of the band renormalization, the electron-phon matrix elements enter as the sum over an infinite number of states, and converged values typically require hundreds of empty states~\cite{ponce_verification_2025}.
That means that for band renormalizations, one should inspect the correction of the electron-phonon matrix elements in a broad range of indexes far from the Fermi level.

Given the smoothness of the corrections that beyond-DFT methods provide on top of DFT results, we introduce an effective scheme to accelerate the convergence for the calculation of the electron-phonon matrix elements with HSE, KI, and $G_0W_0$. 
We define a rescreening factor 
\begin{equation}
 R=\Bigg|\biggl\langle\dfrac{ g_{m n \nu}(\mathbf{k},\mathbf{q})}{ g^{DFT}_{m n \nu}(\mathbf{k},\mathbf{q})}\biggl\rangle\Bigg|^2   
\end{equation}
as the ratio between the beyond-DFT and the DFT electron-phonon matrix elements $g$ averaged over all the bands and phonon indices, and $\mathbf{k}$ and $\mathbf{q}$ points. 
As shown in Table~\ref{tab:rescreen} and in Fig.~\ref{fig:ep_enhancet}, $R$ converges fast with the $\mathbf{k}$ mesh, and can be evaluated from the electron-phonon matrix elements computed on relatively small supercells. 
We finally apply the rescreening factor to the $g^{\rm DFT}$ computed on a dense $\mathbf{k}/\mathbf{q}$ mesh, getting very good estimates of the $g$ for the beyond-DFT methods on the dense grid.
Electronic states labeled by quasi-momentum $\mathbf{k}$ and band index $n$ requires only primitive cell calculations, and are evaluated directly on the same dense grid.
This scheme allows us to avoid prohibitively expensive supercell calculations and, at the same time, obtain converged effective masses (key to getting well-converged transport properties).
Additional details on the rescreening are provided in Figs~\ref{fig:ep_enhancet}, \ref{fig:mob_kcw}, \ref{fig:mob_hse} of the SI.

\subsection{Phonon-limited transport}
Phonon-limited transport is a key example in which electron-phonon interactions play a central role in determining the physical behavior of the system~\cite{Restrepo_2009,claes_phonon-limited_2025}. 
The electric current in semiconductors is related to the electric field $\mathbf{E}$ and the mobility tensor of electrons and holes $\mu_{\alpha \beta}$
\begin{equation}
    \label{eq:mobility}
    \mu_{\alpha \beta} = \sum_{n} \int \dfrac{\mathrm{d}^3 \mathbf{k}}{\Omega^{\rm BZ}} \upsilon_{n \mathbf{k} \alpha} \partial_{E_{\beta}}f_{n \mathbf{k}}.        
\end{equation}
The latter can be computed using a linearized version of the Boltzmann transport equation (BTE)~\cite{ponce_towards_2018} where the derivative of the distribution function $f_{n\mathbf{k}}$ with respect to the external electric field is given by:
\begin{multline}
\label{eq:BTE}
     \partial_{E_{\beta}}f_{n \mathbf{k}} =  e \upsilon_{n \mathbf{k} \beta} \dfrac{\partial f^0_{n \mathbf{k}}}{\partial \varepsilon_{n \mathbf{k}}}\tau_{n \mathbf{k}} \\
    + \dfrac{2 \pi \tau_{n \mathbf{k}}}{\hbar} \sum_{m \nu} \int \dfrac{\mathrm{d}^3 \mathbf{q}}{\Omega^{\rm BZ}} |g_{m n \nu}(\mathbf{k}, \mathbf{q})|^2 [\\(n_{\mathbf{q \nu}} + 1 - f^0_{n \mathbf{k}})\delta(\varepsilon_{n \mathbf{k}} - \varepsilon_{m \mathbf{k}+\mathbf{q}} + \hbar \omega_{\mathbf{q} \nu}) + \\ 
    (n_{\mathbf{q \nu}} + f^0_{n \mathbf{k}})\delta(\varepsilon_{n \mathbf{k}} - \varepsilon_{m \mathbf{k} + \mathbf{q}} - \hbar \omega_{\mathbf{q} \nu})]\partial_{E_{\beta}}f_{m \mathbf{k} + \mathbf{q}},
\end{multline}
 where $\Omega^{\rm BZ}$ is the volume of the first BZ, $\upsilon_{n \mathbf{k}} = \hbar^{-1} \frac{\partial \varepsilon_{n \mathbf{k}}}{\partial \textbf{k}}$, $f^0_{n \mathbf{k}}$ is the Fermi-Dirac occupation function in equilibrium, and $n_{\mathbf{q \nu}}$ is Bose-Einstein distribution. 
The scattering rate $\tau^{-1}_{n \mathbf{k}}$ in Eq.~\eqref{eq:BTE} is
\begin{multline}
    \label{eq:tau}
    \tau^{-1}_{n \mathbf{k}} = \dfrac{2 \pi}{\hbar} \sum_{m \nu} \int \dfrac{\mathrm{d}^3 \mathbf{q}}{\Omega_{BZ}} |g_{m n \nu}(\mathbf{k}, \mathbf{q})|^2 [\\(n_{\mathbf{q \nu}} + 1 - f^0_{m \mathbf{k}+\mathbf{q}})\delta(\varepsilon_{n \mathbf{k}} - \varepsilon_{m \mathbf{k}+\mathbf{q}} - \hbar \omega_{\mathbf{q} \nu}) + \\ (n_{\mathbf{q \nu}} + f^0_{m \mathbf{k}+\mathbf{q}})\delta(\varepsilon_{n \mathbf{k}} - \varepsilon_{m \mathbf{k} + \mathbf{q}} + \hbar \omega_{\mathbf{q} \nu})].
\end{multline}

It has been reported~\cite{Ponce_2021} that the evaluation of mobility with the BTE requires very fine sampling near the edge of the band (typically of the order of 100$^3$ \textbf{k}- and \textbf{q}- grids), which is computationally prohibitive for both DFPT and FD approaches. 
A well established solution is to use Wannier-Fourier interpolations~\cite{marzari_maximally_1997, Giustino_2007}, which allow to exploit the localization of the Wannier functions to perform an accurate interpolation starting from a coarse grid. 
Previous studies show~\cite{ponce_first-principles_2020} that in the case where we do not consider additional scattering mechanisms in Eq.~\eqref{eq:BTE}, the main factors that affect mobility are (i) the curvature of the electronic band structure (effective mass) near the band edges and (ii) the strength of the electron-phonon matrix elements. 
This means that a better description of the electronic band structure, for example using beyond-DFT functionals, could improve the quality of the transport predictions.

\begin{table}[b]
    \caption{\label{tab:table2}%
    Electron and hole drift mobility of silicon and hole mobility of GaAs at 300~K using density-functional theory (DFT), hybrid functional (HSE), Koopman's compliant functional (KI) and $G_0W_0$ using the iterative Boltzmann transport equation.}
    \begin{ruledtabular}
\resizebox{\linewidth}{!}{%
\begin{tabular}{|c|c|c|c|c|c|}
  & \multicolumn{5}{c|}{Drift mobility (cm$^2$/Vs)} \\
  \cline{2-6}
  & DFT & HSE & KI & $G_0W_0$ & Experiment \\
  \hline
   $\mu^{Si}_{h}$ & 517.0 & 486.1 & 483.7 & 461.2 & 445-501~\cite{madelung2002,norton1973, ludwig1956} \\
  \hline
   $\mu^{Si}_{e}$ & 1552.7 & 1393.9 & 1412.3 & 1375.9 & 1350-1450~\cite{jacoboni1977, sze2007} \\
  \hline
   $\mu^{GaAs}_{e}$ & 50005.9 & 12800.8 & 15372.7 & -- & 7200--9750~\cite{madelung2003, hicks1969, rode1971} \\
\end{tabular}%
}
    \end{ruledtabular}
\end{table}

We now investigate the effect of advanced electronic-structure approaches on the transport properties of Si and GaAs. 
We report the electron and hole drift mobilities of silicon as well as electron drift mobilities of GaAs in Table~\ref{tab:table2} using the ab-initio iterative Botlzmann transport equation (BTE)~\cite{ponce_towards_2018,ponce_first-principles_2020} and find that mobility with advanced electronic structure methods is reduced by up to 11\% and 12\% for electrons and holes, respectively, compared to DFT. 
The two main reasons behind this are the effect of rescreening of electron-phonon matrix elements and the change in the curvature of the electronic band structure. 
In the case of Si, these two effects have comparable contributions. 
As seen in Fig.~\ref{fig:el_ph_fine}, the re-screening is higher for the HSE hybrid functional compared to the case of Koopmans. 
However, the resulting mobility value is close to that of Koopmans, which has a smaller rescreening. 
The reason is that for the HSE functional the change in the band structure curvature cancels the rescreening effect on the electron-phonon coupling, resulting in a carrier mobility comparable to that of the Koopmans functional.
In the case of electron mobility with HSE and $G_0W_0$, both increase in the parallel component of the effective mass and electron-phonon coupling lower mobility.
In all cases, we see that accounting for improved electronic description in the electron-phonon couplings improves the predictions of mobilities and brings them closer to the experimental values. 

In the case of GaAs, we find that its electron mobility exhibits a significant change compared to that of DFT when calculated with beyond-DFT approaches.
From Fig.~\ref{fig:mob_gaas}, we see that the effect of advanced functionals is crucial to obtain good agreement with experimental results. 
The main reason is the strong underestimation of the electron effective mass of GaAs in DFT~\cite{dehghani_influence_2025}.
HSE and KI functionals better predict the electron effective mass, and this significantly improves the mobility. 
The secondary effect is the renormalization of electron-phonon couplings, which gives an additional contribution of around 20\% - 30\% with respect to DFT electron-phonon interactions. 
An important conclusion from the calculation of the mobility of GaAs is the significance of band curvature. 
Table~\ref{tab:eff_masses} provides the effective mass on the band-edges for Si and GaAs that are used in the calculations. 
Starting from Si, one can see that the HSE and $G_0 W_0$ are doing well in describing the transversal component of the electron's effective mass if compared to DFT or KI. 
However, they worsen the longitudinal values for electrons and light holes. 
The KI functional does not change significantly the effective masses for this system, and in particular does not modify the hole effective masses. 
This is due to the fact that the 4 $sp^3$ Wannier functions spanning the occupied manifold, and representing the localized manifold of the Koopmans construction, happen to have the same KI correction, leading to a basically rigid shift of the valence manifold without any modification of the dispersion ($\mathbf{k}$-dependence). 
This is not the case for the empty-state manifold, where changes in the band dispersion are observed.
While the rigid shift of the bands is observed here for Si (and similar systems), we stress that it is not a general feature. 
A different choice of the Wannier manifold~\cite{colonna_screening_2018} or the use of more advanced Koopmans flavors, such as KIPZ or pKIPZ, would introduce a non-trivial and \textbf{k}-dependent correction to the DFT bands for both occupied and empty states.
We defer the reader to Fig.~\ref{fig:gaas_kcw_eigs} for more information.

Regardless, both HSE and KI significantly improve the isotropic electron effective mass of GaAs. 
HSE provides improved value for the light-hole effective mass, but for accurate transport calculation, it is necessary to include spin-orbit coupling. 
Overall, and as a general conclusion, note that the description of electron mobility benefits from the usage of beyond-DFT functionals.      
\begin{table*}
    \centering
    \caption{Effective masses for Si and GaAs with different methods (in units of the free electron mass, \(m_e\)).}
    \label{tab:eff_masses}
    \begin{tabular}{|l|l|c|c|c|c|}
        \hline
        Material & Method & Electron $m^*_{||}$ & Electron $m^*_\perp$ & Hole $m^*_{hh}$ & Hole $m^*_{lh}$ \\
        \hline
        \multirow{4}{*}{Si} 
        & DFT        & 0.859 & 0.202 & 0.276 & 0.177 \\
        & HSE        & 1.099 & 0.189 & 0.276 & 0.185 \\
        & KI         & 0.820 & 0.208 & 0.276 & 0.177 \\
        & $G_0 W_0$  & 0.905 & 0.198 & 0.284 & 0.188 \\
        & Experiment & 0.920 & 0.190 & 0.490 & 0.160 \\
        \hline
        \multirow{4}{*}{GaAs} 
        & DFT        & \multicolumn{2}{c|}{0.027} & 0.316 & 0.027 \\
        & HSE        & \multicolumn{2}{c|}{0.058} & 0.304 & 0.057 \\
        & KI         & \multicolumn{2}{c|}{0.054} & 0.316 & 0.027 \\
        & Experiment & \multicolumn{2}{c|}{0.067} & 0.510 & 0.080 \\
        \hline
    \end{tabular}
\end{table*}

\section{Discussion}
In this work, we introduced a novel framework to calculate electron-phonon couplings that works with any electronic-structure method, as long as it can provide electronic eigenvalues and eigenvectors of perturbed and unperturbed systems. 
The approach is general, allows integration with many modern DFT and beyond-DFT methods, and is elegant in bypassing the challenges of direct methods (beyond norm-conserving pseudopotentials and orbital-dependent operators) through a novel projectability approach.

Using symmetry considerations, we devised an efficient algorithm to unfold wave functions to all required supercells, drastically decreasing the number of calculations needed to perform. 
The whole approach is implemented in a new dedicated electron-phonon code named \textsc{ElePhAny} that is interfaced with the \textsc{Quantum ESPRESSO} and the \textsc{Koopmans} packages as well as \textsc{Yambo} and allows the evaluation of electron-phonon coupling for all functionals available in the mentioned codes. 
After that, an interface to the \textsc{EPW} code allows the calculation of any electron-phonon-related properties that require fine sampling of the Brillouin zone.

We used the proposed approach to study the electron-phonon coupling, effective masses, and mobilities in Si and GaAs. 
Using advanced electronic structure methods leads to a significantly improved description of electron mobilities in GaAs and results in a closer agreement with experimental results.
The proposed approach paves the way for the calculation of electron-phonon-related properties with a realistic electron description.

\section{Methods}
\subsection{Computational workflow}
\begin{figure*}
    \includegraphics[width=1.0\linewidth]{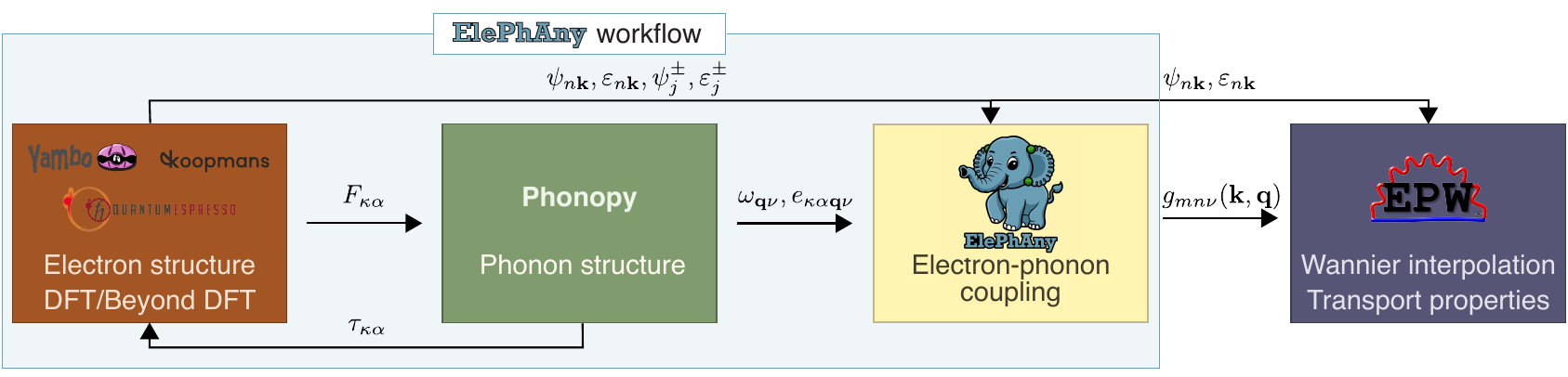}
    \caption{\label{fig:workflow}
    Schematic representation of the workflow. 
    The \textsc{ElePhAny} code serves as the main driver for orchestrating the preceding calculations of electronic structure and phonon properties, as well as for calculating electron-phonon matrix elements using the projectability approach. 
    As the initial step, the code calls the \textsc{Phonopy} application programming interface to create displaced supercell configurations based on the input of the preferred electronic structure code (\textsc{Quantum ESPRESSO}, \textsc{Koopmans}, \textsc{Yambo}).
    After running the chosen electronic structure code and providing a set of resulting wave functions (in binary or \texttt{hdf5} format) and eigenvalues in \textsc{XML} format together with Phonopy output of phonon eigendisplacements and frequencies in \textsc{YAML} format, the ElePhAny code constructs an electron-phonon matrix and saves it in the binary \textsc{EPW} \texttt{.epb} format for further calculation of electron-phonon related properties.
    }
\end{figure*}
Calculating transport properties using the proposed approach requires several steps. 
Starting with the atomic structure, one (i) performs self-consistent calculations for the pristine and perturbed systems, 
(ii) calculates phonon dispersion, (iii) computes electron-phonon matrix elements on a coarse grid, (iv) Wannierizes target manifold and interpolate all relevant quantities on a fine grid, and (v) solves the Boltzmann transport equation with the \textsc{EPW} code. 
Some of these tasks are achieved by using existing codes. 
In particular, we use \textsc{Quantum ESPRESSO}~\cite{Giannozzi_2009, Giannozzi_2017}, \textsc{Koopmans}~\cite{linscott_koopmans_2023} and \textsc{Yambo}~\cite{marini_yambo_2009} for DFT, hybrid, Koopmans and $G_0W_0$  calculations, \textsc{Phonopy}~\cite{phonopy_2015} for the generation of supercells with displaced atoms and phonon dispersion, and \textsc{EPW}~\cite{ponce_epw_2016,lee_electronphonon_2023} for Wannier interpolation and BTE solution. 
For the remaining tasks, we developed a flexible and modular \textsc{Julia} code called \textsc{ElePhAny} that evaluates the electron-phonon matrix elements on a coarse grid with the projectability approach, and interfaces and manages the other software involved to obtain all the quantities required for the electron-phonon matrix elements calculations. 
The overall workflow is depicted on Fig.~\ref{fig:workflow}.  
Although we used the \textsc{Quantum ESPRESSO} package,
we stress that the approach is general and can be applied to any first-principles software that provides eigenvalues and forces. 
The electron-phonon matrix elements in the Kohn-Sham (KS) basis are saved in a binary format readable by \textsc{EPW} that is used for the subsequent Wannier interpolation and transport properties calculation. 
A \textsc{Phonopy} interface is used to generate the inputs for the perturbed supercell calculations.
Since phonon properties could also be computed using only inequivalent displacements, the \textsc{ElePhAny} code checks whether the proposed displacements would allow to obtain all perturbed configurations using Eq.~\eqref{eq:symmetries}. 
After having performed supercell calculations, the most time-consuming part of the evaluation of the electron-phonon matrix element via the projectability approach is the evaluation of Eq.~\eqref{eq:projectability} for each \textbf{k} and \textbf{q} points. 
Although we do not exploit it here, we note that this could be efficiently parallelized over \textbf{k} points.
\subsection{Computational details}
To demonstrate the approach, we performed electron-phonon and mobility calculations for Si and GaAs using PBE, HSE and KI functionals.
For both systems, we used supercells of sizes 2$\times$2$\times$2, 3$\times$3$\times$3, and 4$\times$4$\times$4 to demonstrate the convergence of rescreening of the electron-phonon matrix elements. 
In both calculations, the hybrid functional was chosen as HSE06~\cite{Heyd2003,Heyd2006} with a fraction of the exact exchange set to the default of 0.25.
For Si, we also performed $G_0W_0$ calculations in the pristine unit cell, as well as in 2$\times$2$\times$2 and 4$\times$4$\times$4 supercells with displaced atoms. 
The finite displacement step was chosen to be 0.001~Bohr to ensure convergence of the FD approach. 

For Si, we used a optimized norm-conserving Vanderbilt pseudopotential\cite{Hamann_2013,hamann_erratum_2017} from the DOJO library~\cite{van_setten_pseudodojo_2018} wavefunctions cutoff of 60~Ry. 
For Wannier interpolation of band structure and electron-phonon matrix elements, we used sp$^3$ orbitals as initial projections for occupied bands and $s+d$ for empty ones. 
To perform the disentanglement procedure, we calculated 20 bands. 
In the case of the Koopmans KI functional, the same initial projections were chosen as variational orbitals for primitive cell calculations. 
From the tests we observed that even though the effective mass is significantly affected by the quality of interpolation, the electron-phonon coupling with the Koopmans functional was much less sensitive to that.
For that reason, supercell calculations were performed with sp$^3$ orbitals as initial projections for empty bands to reduce the computational cost to 4 empty orbitals. 
The screening $\alpha$ parameters were calculated with a linear response scheme~\cite{colonna_screening_2018, colonna_koopmans_2022} on the 6$\times$6$\times$6 grid and fixed for the Hamiltonian calculation with finer grids. 
For the final mobility calculation, the coarse grid of 12$^3$ ensures convergence of effective mass for electrons. 
Such a large coarse grid was required since the conduction band minima do not lie on a high symmetry point~\cite{Ponce_2021}. 
For the fine grids, values of 60$^3$ and 100$^3$ were chosen.  
The $G_0W_0$ calculation was performed using 100 states of the corresponding pristine unit cell (6400 states in the 4$\times$4$\times$4 supercell). 
For the dielectric function evaluation we rely on the plasmon pole approximation \cite{aryasetiawan_GW_1998} and use G-vectors cutoff of 8~Ry for dielectric matrix.

For GaAs, a Schlipf-Gygi optimized norm-conserving Vanderbilt pseudopotential ~\cite{schlipf_optimization_2015} with wavefunctions cutoff 80~Ry was used.
However, for the HSE functional, we had to use the cutoff of 100~Ry to overcome the issue of soft modes in phonon calculations with hybrid functional. 
For both occupied and empty bands, sp$^3$ orbitals were used. 
Moreover, the wannierization was performed by mixing occupied and empty bands to improve the localization by increasing the number of degrees of freedom in the minimization procedure. 
KI functional was also built on top of the Wannier functions with sp$^3$ orbitals as initial projections. 
The screening parameters were calculated on the 4$\times$4$\times$4 grid and then fixed for further Hamiltonian calculations. 
In this case, the orbitals are not mixed, since the Koopmans functional tends to favor block-by-block wannierization~\cite{de_gennaro_blochs_2022}. 
As the final coarse grid for obtaining eigenvalues of the Hamiltonian and rescreened electron-phonon coupling, the 8$^3$ sampling and 60$^3$-350$^3$ fine grid were chosen.   

\section*{Data availability}
The data needed to reproduce the results in this work can be found in the Materials Cloud Archive~\cite{poliukhin_2025_pt38k-stk16}.

\section*{Code availability}
The source code is available on GitHub (https://github.com/Koulb/ElePhAny.jl).

\section*{Acknowledgements}
A.P. thanks Miki Bonacci, Matteo Quinzi, Riccardo De Gennaro, and Gabriel de Miranda Nascimento for fruitful discussions about GW calculations, derivatives of non-local parts of pseudopotentials, Koopmans functionals, and symmetries of displaced calculations, respectively.
A.P. also thanks Ekaterina Poliukhina for the help with graphics in the manuscript.
The authors acknowledge support from the Swiss Platform for Advanced Scientific Computing, project: ``Spectral properties of materials on accelerated architecture'' and the Swiss National Science Foundation, grant number 213082: ``Accurate and efficient electronic structure functionals for energies and spectra of materials''.
S. P. is a Research Associate of the Fonds de la Recherche Scientifique - FNRS.
This work was supported by the Fonds de la Recherche Scientifique - FNRS under Grants number T.0183.23 (PDR) and  T.W011.23 (PDR-WEAVE). This publication was supported by the Walloon Region in the strategic axe FRFS-WEL-T.
N.M. acknowledges support by the NCCR MARVEL, a National Centres of Competence in Research, funded by the Swiss National Science Foundation (grant number 205602)
The computational time has been provided by the Swiss National Supercomputing Centre (CSCS) under project ID mr33 and the IMX cluster at EPFL.

\section*{Author contributions}
A.P. proposed the method, designed and developed the code, wrote the first draft of the manuscript, and ran and analyzed the calculations.
N.C. contributed to running the Koopmans calculations for primitive cells and supercells, discussed the results, and contributed to the draft and final version of the manuscript.
F.L. wrote the initial version of the code based on potential differences and contributed to the final version of the manuscript.
S.P. contributed to the execution of EPW-related calculations, discussed the results, and contributed to the draft and final version of the manuscript.
N.M. supervised the project, discussed the results, and contributed to the final version of the manuscript.

\section*{Competing interests}
The authors declare no competing interests.
%
\putbib[main]
\end{bibunit}

\clearpage

\setcounter{equation}{0}
\renewcommand{\theequation}{S\arabic{equation}}
\setcounter{figure}{0}
\renewcommand{\thefigure}{S\arabic{figure}}
\setcounter{table}{0}
\renewcommand{\thetable}{S\arabic{table}}
\setcounter{affil}{0}

\begin{bibunit}[apsrev4-2]

\title{Supplementary Information:\\
Carrier mobilities and electron-phonon interactions beyond DFT}

\author{Aleksandr Poliukhin}
\affiliation{Theory and Simulation of Materials (THEOS), École polytechnique fédérale de Lausanne, 1015 Lausanne, Switzerland}
\email{aleksandr.poliukhin@epfl.ch}
\author{Nicola Colonna}%
\affiliation{PSI Center for Scientific Computing, Theory and Data, 5232 Villigen PSI, Switzerland}%
\author{Francesco Libbi}
\affiliation{John A. Paulson School of Engineering and Applied Sciences, Harvard University, Cambridge, MA 02138, USA}%
\author{Samuel Poncé}
\affiliation{European Theoretical Spectroscopy Facility, Institute of Condensed Matter and Nanosciences, Université catholique de Louvain, Chemin des Étoiles 8, B-1348 Louvain-la-Neuve, Belgium}%
\affiliation{WEL Research Institute, avenue Pasteur 6, 1300 Wavre, Belgium}
\author{Nicola Marzari}
\affiliation{Theory and Simulation of Materials (THEOS), École polytechnique fédérale de Lausanne, 1015 Lausanne, Switzerland}
\affiliation{PSI Center for Scientific Computing, Theory and Data, 5232 Villigen PSI, Switzerland}%
\date{\today}

\maketitle
\onecolumngrid

\section*{Finite difference approach with derivatives of the potential}
In a conventional finite difference approach, the derivatives that are required in Eq.~\eqref{eq:g_full} could be calculated as the finite difference of the effective potential:
\begin{equation}
    \label{eq:pot_dif}
    \dfrac{\partial V}{\partial \tau_{l\alpha}}(\mathbf{r}) \approx \frac{V^{+}(\mathbf{r}) - V^{-}(\mathbf{r})}{\tau_{l\alpha}},
\end{equation}
\noindent where $V^{\pm}$ are effective KS potentials of perturbed supercell. 
The caveat in such an approach is that the total potential often consists of local and non-local parts, where the latter is due to standard pseudopotential construction~\cite{kleinman_efficacious_1982,van_setten_pseudodojo_2018}. 
Therefore, an individual treatment is required for different types of pseudopotentials. 
Recent works in this direction show that the expression for PAW pseudopotentials can be challenging to implement~\cite{chaput_finite-displacement_2019}. 
In the present work, we considered only norm-conserving pseudopotential, for which we derived the following contribution of the non-local part of the potential to the corresponding brakets:
\begin{align}
    \label{eq:elph:NL}
    \bra{\psi_{i \textbf{k}+\textbf{q}}}\dfrac{\partial V^{\rm NL}}{\partial \textbf{R}}\ket{\psi_{j \textbf{k}}} =&  \sum_{IJ} D_{IJ} \Bigg[\left(\dfrac{\partial}{\partial \textbf{R}}\bra{\psi_{i \textbf{k}+\textbf{q}}}\ket{Y\beta_I}\right)\bra{Y\beta_J}\ket{\psi_{j \textbf{k}}} +  \bra{\psi_{i \textbf{k}+\textbf{q}}}\ket{Y\beta_I}\left(\dfrac{\partial}{\partial \textbf{R}}\bra{Y\beta_J}\ket{\psi_{j \textbf{k}}}\right) \Bigg], \\
    \dfrac{\partial}{\partial \textbf{R}}\bra{Y\beta_I}\ket{\psi_{j \textbf{k}}} =& \sum_{\textbf{k}+\textbf{G}}\bra{Y\beta_I}i(\textbf{k}+\textbf{G})e^{i(\textbf{k}+\textbf{G})\cdot\textbf{R}}\ket{\textbf{k}+\textbf{G}}\bra{\textbf{k}+\textbf{G}}\ket{\psi_{j \textbf{k}}}, \\
    \bra{\textbf{G}}\ket{Y\beta_{I={\{l,m,\varepsilon\}}}} =& \dfrac{4 \pi}{\sqrt{V}}Y_{lm}\left(\frac{\textbf{G}}{|\textbf{G}|}\right)\int_{0}^{\infty}r^2 J_{l}(|\textbf{G}|r)\beta_{\varepsilon}(r)\mathbf{d}r,
\end{align}
where $V^{\rm NL} = \sum_{IJ} D_{IJ}\ket{Y\beta_I}\bra{Y\beta_J}$ with $Y\beta_I(\mathbf{r}) = Y_{lm}(\mathbf{r})\cdot \beta_\varepsilon(r) $ is written as an expansion on spherical harmonics $Y_{lm}$ that have angular momentum $l$ and magnetic quantum number $m$ with the short-ranged radial functions $\beta_{\varepsilon}(r)$ constructed using reference energy $\varepsilon$ to match logarithmic derivative of the all-electron wavefunction, combined index were set to $I = \{l,m,\varepsilon\}$, $J = \{l',m',\varepsilon'\}$, $D_{IJ}$ are non-local pseudopotential projector coefficients, $J_{l}$ is the $l$-order spherical Bessel functions.

\begin{figure}[h]
    \includegraphics[width=0.85\columnwidth]{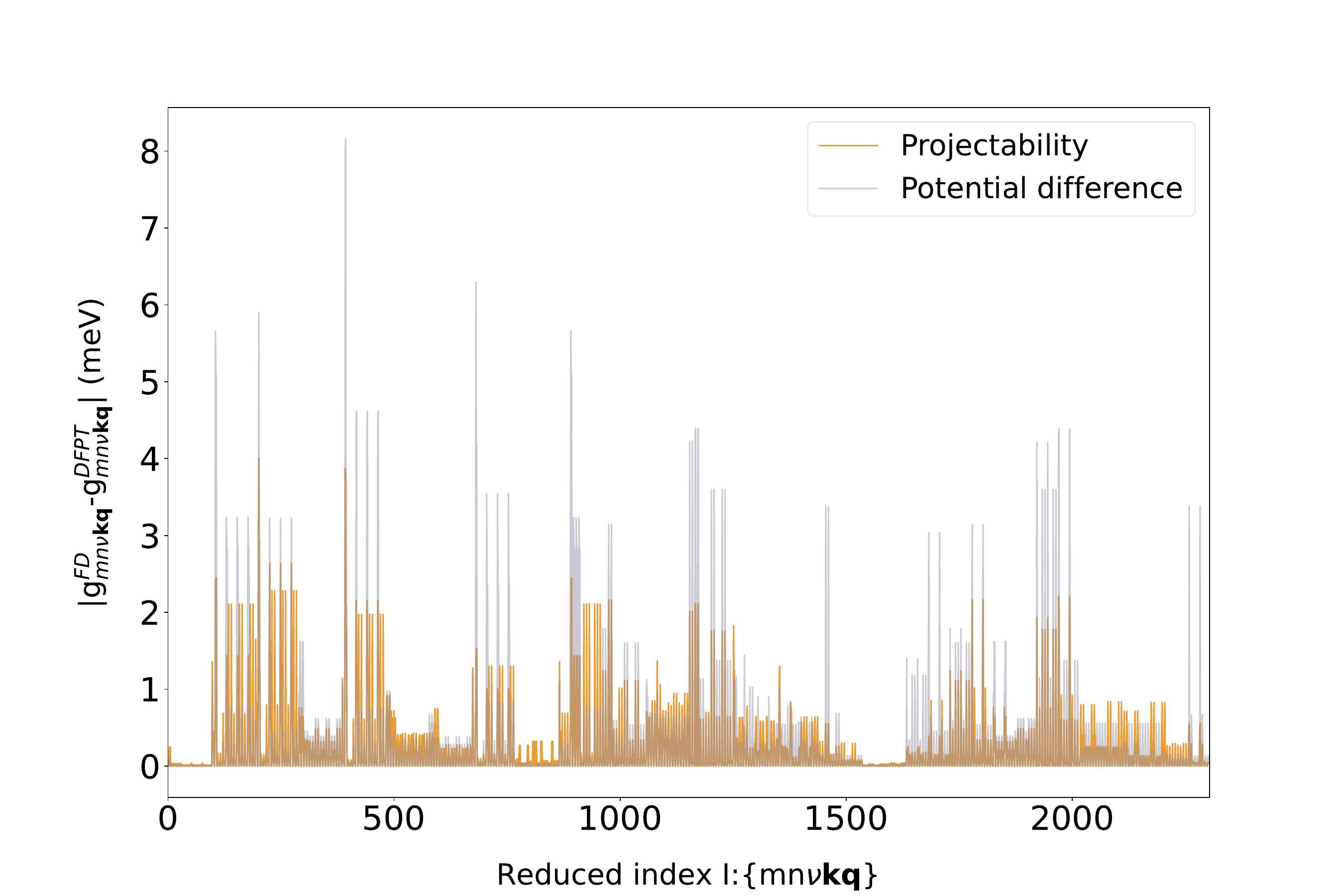}
    \caption{ \label{fig:pot_vs_proj}Comparison of 2 finite-difference (FD) approaches: the projectability approach described in the paper and the conventional potential differences for a 2$\times$2$\times$2 supercell of Si with absolute displacement of 10$^{-2}$ Bohr. 
    The error of both methods lies in the same range, even though the projectability approach is more accurate for some values.}
\end{figure}

Using \eqref{eq:elph:NL} together with the local contribution of the KS potential, one can obtain electron-phonon matrix elements at arbitrary rational \textbf{k} and \textbf{q} points when the commensurate supercell is used. 
Fig.~\ref{fig:pot_vs_proj} shows that the projectability and finite difference (FD) of the potentials approaches give a comparable accuracy compared to DFPT.

\newpage
\clearpage
\section*{Convergence of finite difference approach}
\begin{figure*}[ht]
    \includegraphics[width=0.8\textwidth]{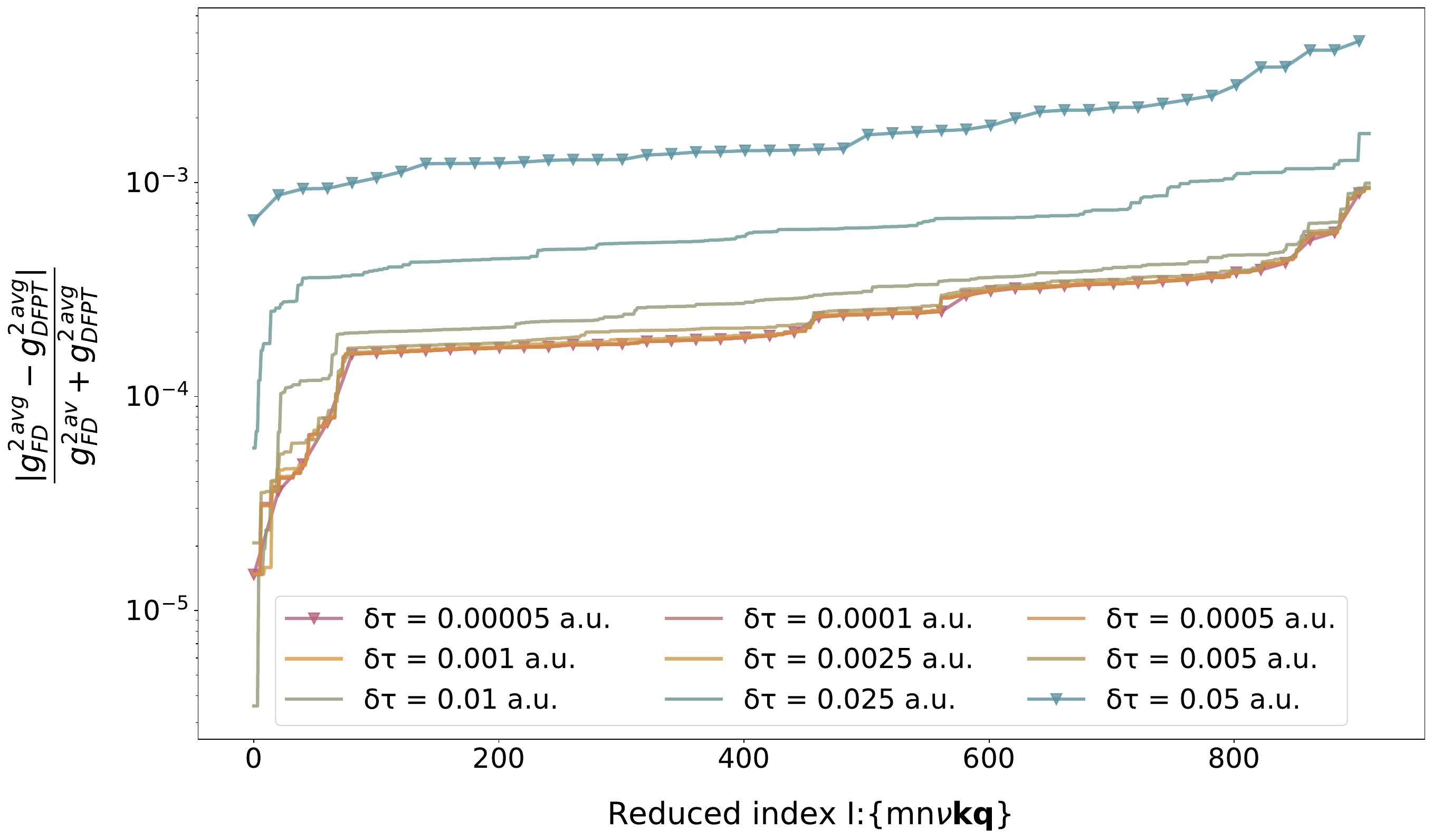}
    \caption{\label{fig:dfpt_converge}Relative convergence of FD approach with respect to finite difference step for the 2$\times$2$\times$2 \textbf{q}-grid of silicon. 
    It is seen that the finite step below or equal to 10$^{-3}$~Bohr allows the convergence of most of the electron-phonon matrix elements to the relative error of 10$^{-3}$--10$^{-4}$.}
\end{figure*}

\begin{figure*}[ht]
    \includegraphics[width=0.8\textwidth]{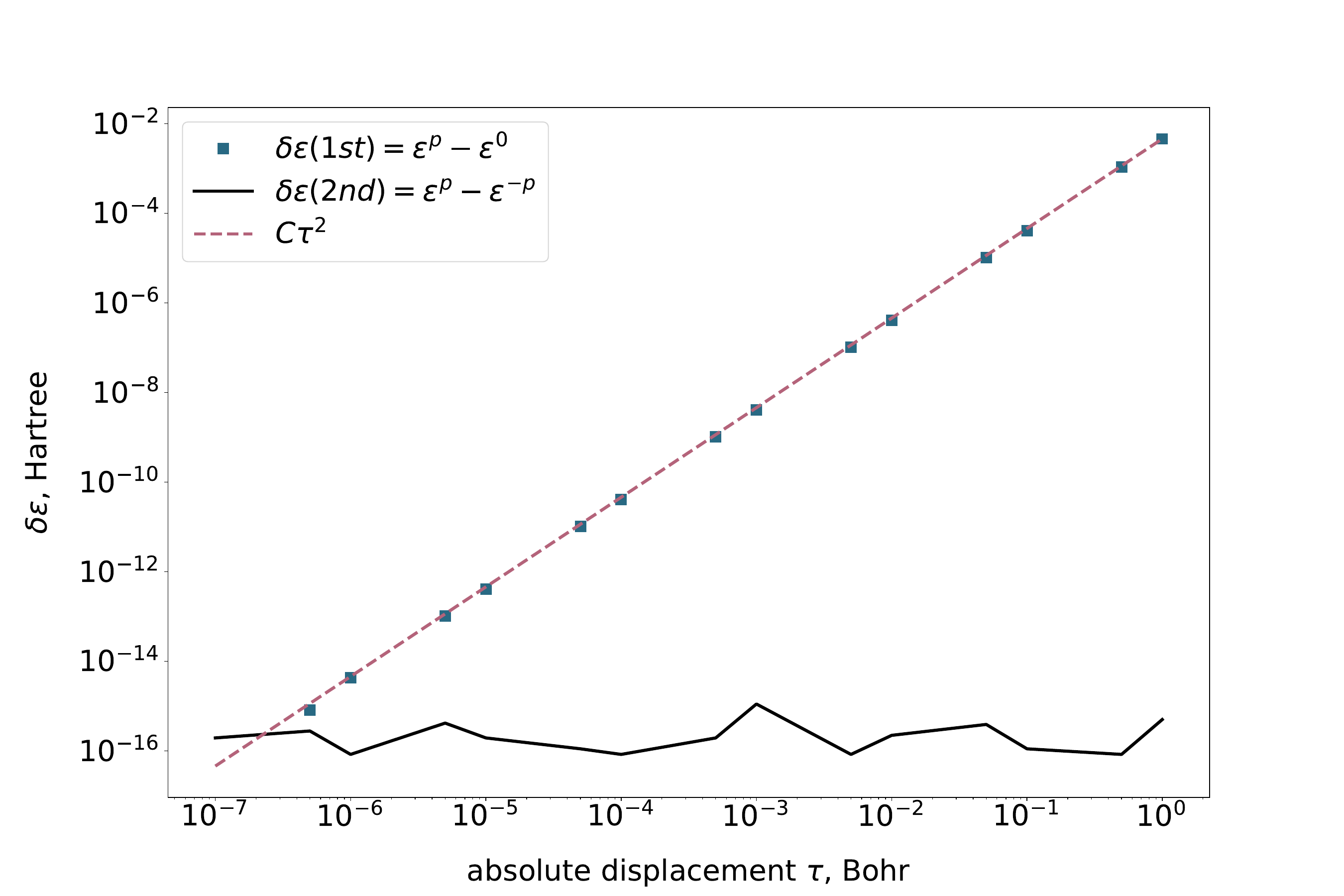}
    \caption{\label{fig:converge_1ep}Difference between perturbed eigenvalues computed with different finite difference schema for the first energy level of silicon at \textbf{k}=$\Gamma$. 
    The electron-phonon matrix element should be zero for this band so it represents the convergence of the finite difference approach in this specific case.}
\end{figure*}

To ensure the reliability of our calculations, we performed various convergence tests. 
Fig.~\ref{fig:dfpt_converge} illustrates the typical convergence behavior of electron-phonon matrix elements as a function of the finite difference step size. 
The results show that convergence is achieved for step sizes smaller than $\delta \tau < 10^{-2}$ Bohr. 
Beyond this point, the convergence reaches a plateau and does not improve any further. 
This plateau for FD methods is well known and stems from numerical accuracies~\cite{phonopy_2015,LaflammeJanssen2016}.
This can introduce minor inconsistencies between FD and DFTP that prevent the exact numerical agreement of the electron-phonon matrix.  
However, the FD projectability formula in Eq.~\eqref{eq:projectability} proposed in this work can show a convergence trend approaching numerical precision. 
In the case of the non-degenerate band at \textbf{q} = $\Gamma$, Eq.~\eqref{eq:projectability}, reduces to the eigenvalue differences divided by the absolute displacement. 
By examining the specific case where DPFT predicts this braket to go to zero, we can analyze the convergence of the FD approach. 
This is shown in Fig.~\ref{fig:converge_1ep} where we examine the convergence of the eigenvalue difference for a selected silicon band ($m$ = $n$ = 1, \textbf{k} = $\Gamma$). 
We see that the eigenvalue difference goes to zero faster than the displacement step as the latter goes to zero.
Additionally, the effect of using different discretization schemes is also presented in Fig.~\ref{fig:converge_1ep}.
When first-order FD is used, we need to go to minimal displacements to converge whereas second-order FD delivers numerical accuracy for all the range of FD considered. 
We remark that we did not use crystal symmetries for these calculations since \textsc{Quantum ESPRESSO} would 
detect them to be the same when the displacements are smaller than 10$^{-5}$~Bohr.

\begin{figure*}[ht]
    \includegraphics[width=0.7\textwidth]{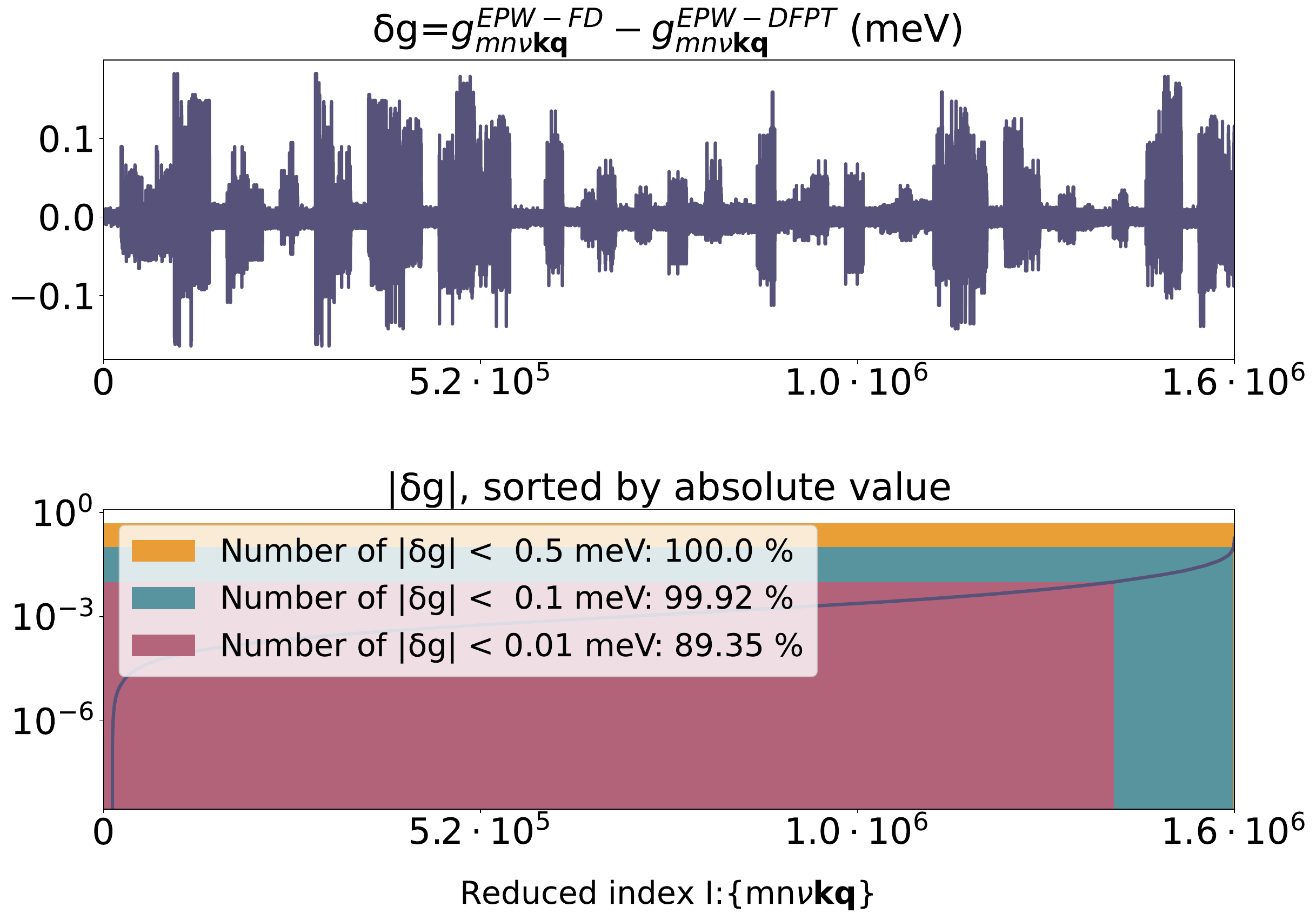}
    \caption{\label{fig:epw_converge}The absolute error of matrix elements produced by EPW, using FD and DFPT as an initial method for 4$\times$4$\times$4 \textbf{q}-grid of GaAs. 
    Most of the electron-phonon matrix elements differ by less than 0.01~meV, which allows for the correct reproduction of results for mobilities on the DFT level.}
\end{figure*}

To further assess the accuracy of our FD approach compared to DFPT, it is also important to check the consistency between electron-phonon matrix elements produced by \textsc{EPW} after performing the Wannier interpolation when using the conventional DFPT or the FD interface proposed in this work. 
This is illustrated in Fig.~\ref{fig:epw_converge}.
Most of the electron-phonon matrix elements agree with an error of less than 0.01~meV, which allows to predict the same DFT mobility when using both schemes.


\newpage
\clearpage
\section*{Applicability of the projectability approach}
In order to understand the applicability of the projectability approach for the calculation of the electron-phonon matrix element, we can estimate the error that is made when a finite number of bands is used in Eq.~\eqref{eq:projectability}. 
We know that electron-phonon matrix elements correspond to the first order term in the expansion of the potential energy with respect to the atomic displacement ($ g_{mn\nu} =  \mathcal{O}(\tau)$). 
Without loss of generality, we consider the case of \textbf{k}=\textbf{q}=$\Gamma$:
\begin{equation*}
    g_{mn\nu } \propto \bra{\psi_m} \dfrac{\partial \hat{H}}{\partial \tau_{
    \kappa\alpha}} \ket{\psi_n}\tau \Rightarrow \bra{\psi_m} \dfrac{\partial \hat{H}}{\partial \tau_{\kappa\alpha}}\ket{\psi_n} = \mathcal{O}(1).   
\end{equation*}

Now, we introduce a finite-size basis approximation where in Eq.~\eqref{eq:projectability} we truncate the sum by only considering $N^{\rm max}$ states. 
In the derivation below, we consider that states $m$ and $n$ of the electron-phonon matrix have ``corresponding perturbed states'' $m'$ and $n'$ among the first $N^{\rm max}$ states. 
Those are the states that are adiabatially connected to the unperturbed ones when the atoms are slightly displaced from their equilibrium positions.
In this approximation, we neglect terms with a quantum number greater than $N^{\rm max}$, and the matrix elements of the change in the Hamiltonian can be written in the following way:
\begin{align}
        \bra{\psi_m} \dfrac{\partial \hat{H}}{\partial \tau_{\kappa\alpha}} \ket{\psi_n} =&  \frac{\tau^{-1}_{\kappa\alpha}}{2}\sum_{j} (\varepsilon_j^{+} u^{*+}_{mj}u^{+}_{jn}  - \varepsilon_j^{-} u^{*-}_{mj}u^{-}_{jn} )  + \mathcal{O}(\tau^2)  \\
        =&   \frac{\tau^{-1}_{\kappa\alpha}}{2}\sum^{j \le N^{\rm max}}_{j} (\varepsilon_j^{+} u^{*+}_{mj}u^{+}_{jn}  - \varepsilon_j^{-} u^{*-}_{mj}u^{-}_{jn} ) +   \frac{\tau^{-1}_{\kappa\alpha}}{2}\sum^{j>N^{\rm max}}_{j}( \varepsilon_j^{+} u^{*+}_{mj}u^{+}_{jn}  - \varepsilon_j^{-} u^{*-}_{mj}u^{-}_{jn} )   + \mathcal{O}(\tau^2) \\
        =&  \frac{\tau^{-1}_{\kappa\alpha}}{2}\sum^{j \le N^{max}}_{j} (\varepsilon_j^{+} u^{*+}_{mj}u^{+}_{jn}  - \varepsilon_j^{-} u^{*-}_{mj}u^{-}_{jn} ) + \Delta^{+}_{>N^{max}} + \Delta^{-}_{>N^{\rm max}}   + \mathcal{O}(\tau^2)
\end{align}
where $\Delta^{\pm}_{>N}$ includes the contribution to the sum of states larger than N$^{\rm max}$. 
By analogy, we can define the $\Delta^{\pm }_{\le N}$ where the sum is instead restricted to the first $N$ states. 
If we consider the case of non-degenerate states and using for perturbed states first-order perturbation theory, one arrives at the following expressions:
\begin{align}
    \label{eq:project_err1}
     \Delta^{+}_{>N^{\rm max}} =& \frac{\tau^{-1}_{\kappa,\alpha}}{2}\sum^{j > N^{\rm max}}_{j} \varepsilon_j^{+} u^{*+}_{mj}u^{+}_{jn}  =  \frac{\tau^{-1}_{\kappa\alpha}}{2} \sum^{j > N^{\rm max}}_{j}  (\varepsilon_j + \Delta V_{jj} \tau_{\kappa\alpha}) \frac{\Delta V^{*}_{mj}\tau_{\kappa\alpha}}{\varepsilon_j - \varepsilon_m}\frac{\Delta V_{jn}\tau_{\kappa\alpha}}{\varepsilon_j - \varepsilon_n} + \mathcal{O}(\tau^2) =  \mathcal{O}(\tau) \\
     \Delta^{+}_{\le N^{\rm max}}  =& \frac{\tau^{-1}_{\kappa\alpha}}{2}\sum^{j \le N^{\rm max}}_{j} \varepsilon_j^{+} u^{*+}_{mj}u^{+}_{jn} =  \frac{\tau^{-1}_{\kappa\alpha}}{2} \sum^{j \le N^{\rm max}}_{j}  (\varepsilon_j + \Delta V_{jj} \tau_{\kappa\alpha}) \delta_{mj} \delta_{jn} + \mathcal{O}(\tau^2) =  \mathcal{O}(\tau^{-1}), \label{eq:project_err2}
\end{align}
where $\Delta V_{jj} = \bra{\psi_j}\frac{\partial V}{\partial \tau_{\kappa, \alpha}}\ket{\psi_j}$. 
Eqs.~\eqref{eq:project_err1} and \eqref{eq:project_err2} represent the estimation of an error that is made when not accounting for the corresponding term. 
From this equation, we conclude that omitting the term Eq.~\eqref{eq:project_err2} or equivalently calculating the electron-phonon matrix elements for unperturbed states $m$ or $n$ greater than $N^{\rm max}$, which means that for states that do not have corresponding perturbed states, introduces an error of $\tau^{-1}$, which makes the approximation inadequate. 
Thus, if one wants to describe electron-phonon matrix elements up to the electronic state $N^{\rm max}$, all the corresponding perturbed states with a number smaller than or equal to $N^{\rm max}$ should be included in the sum. 
The only exception is the degenerate states. 
If the last state $N$ is degenerate and only the part of degenerate manifold is included in the sum, the error would be of the same order as Eq.~\eqref{eq:project_err2}. 
This is shown in Fig.~\ref{fig:projectability_dft} where the first 20 states of Si are considered. 
The last two states are triply degenerate, and to ensure the condition in Eq.~\eqref{eq:projectability_condition}, we include one more state in the sum. 
The general solution in the computation is always to check Eq.~\eqref{eq:projectability_condition} and include more states if it is violated. 

\newpage
\clearpage

\section*{Electron-phonon matrix elements with beyond-DFT methods}
\begin{figure}[ht]
    \centering
    \includegraphics[width=0.65\textwidth]{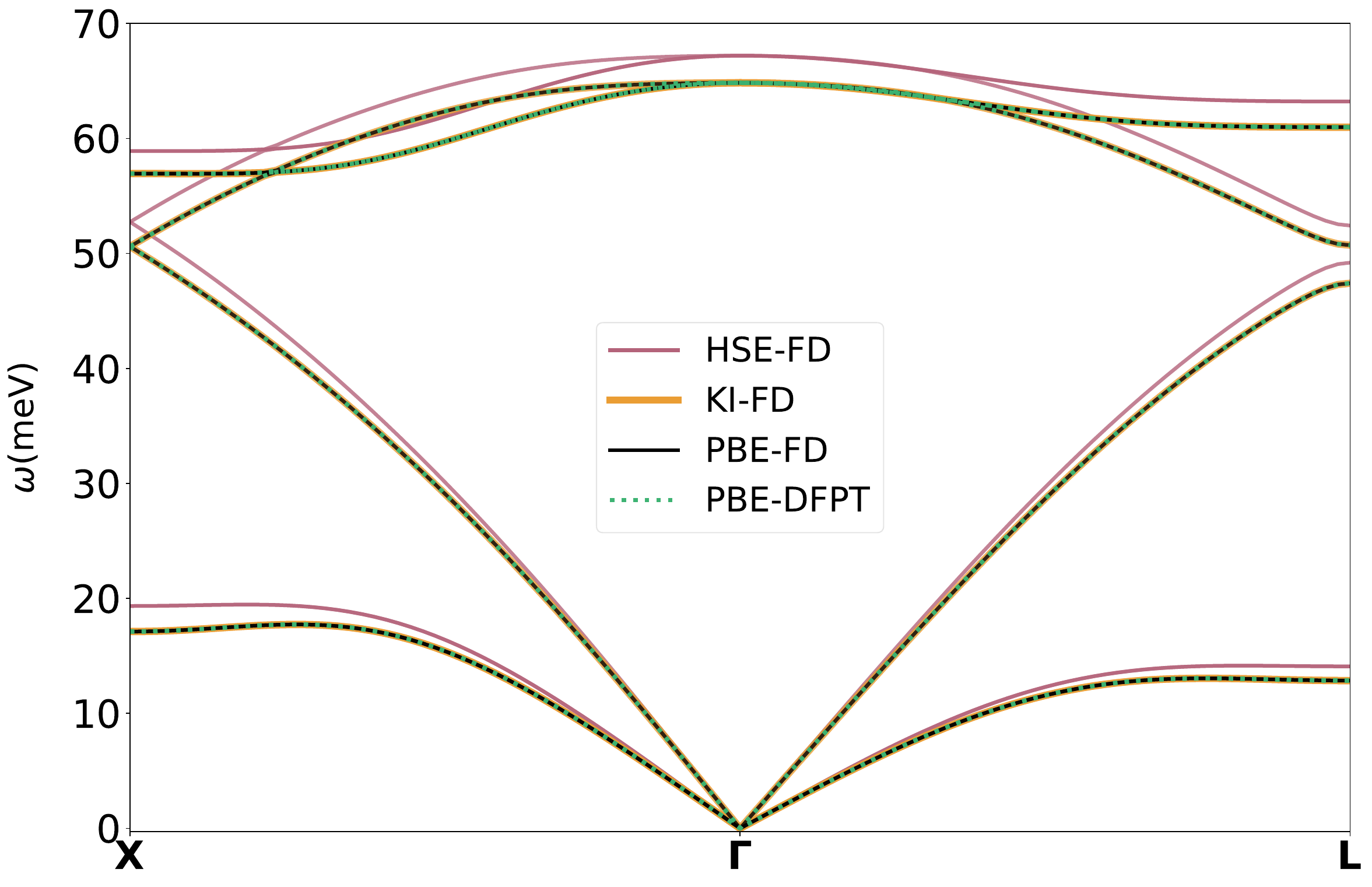}
    \caption{\label{fig:ph_epw} Phonon band structure of Si with different functionals. 
    The interpolation is performed from a 4$\times$4$\times$4 supercell.}
\end{figure}

\begin{figure}[ht]
    \includegraphics[width=0.65\linewidth]{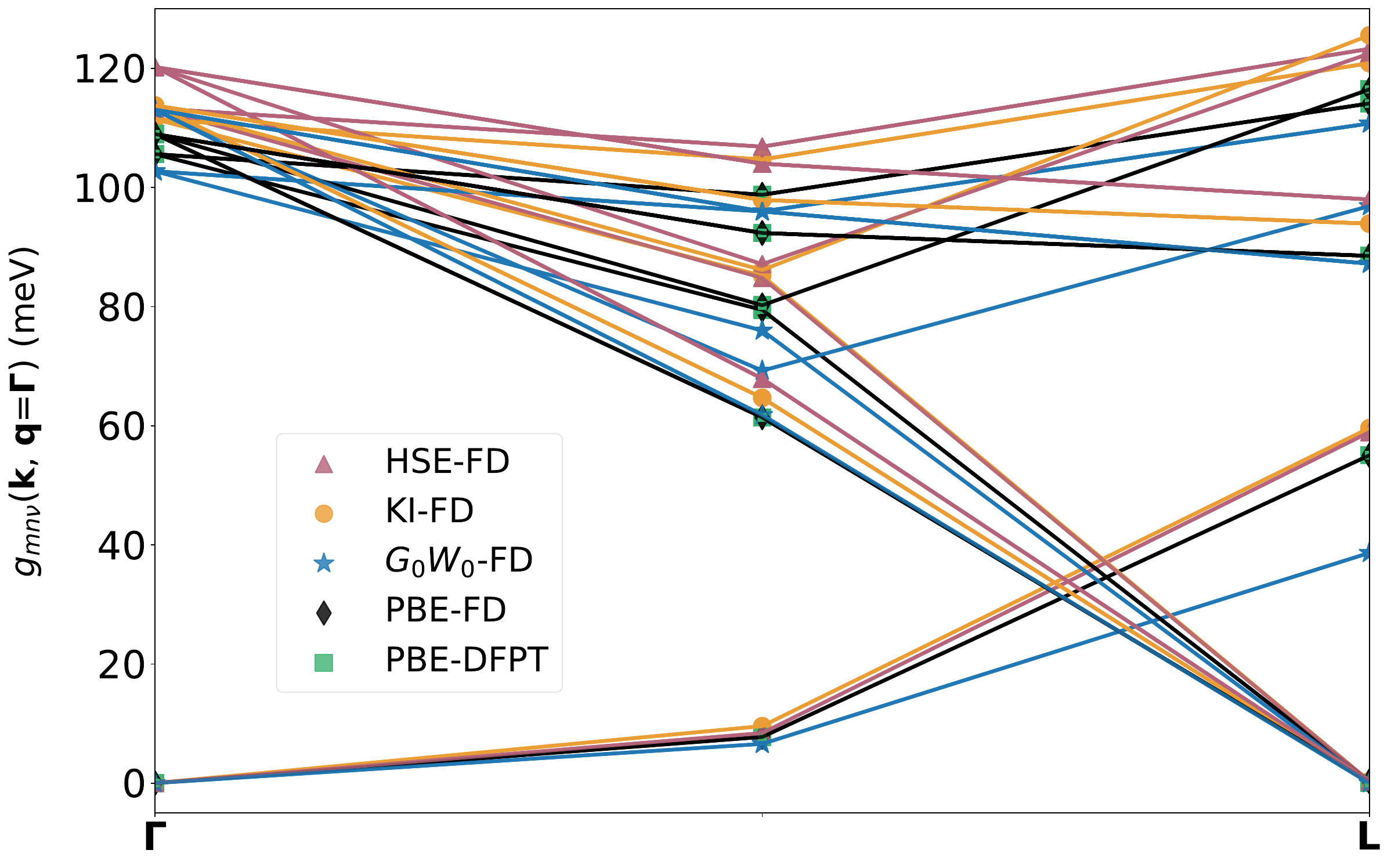}
    \caption{\label{fig:el_ph_coarse}
    Electron-phonon matrix elements of Si with different functionals for the first 4 occupied bands for \textbf{q}=$\boldsymbol{\Gamma}$ and \textbf{k} along the high symmetry lines $\boldsymbol{\Gamma}-\boldsymbol{L}$.
    HSE and KI functionals yield an uniform increase in the values of the electron-phonon matrix elements compared to DFT.}
\end{figure}

\begin{figure}[ht]
    \centering
    \includegraphics[width=0.7\textwidth]{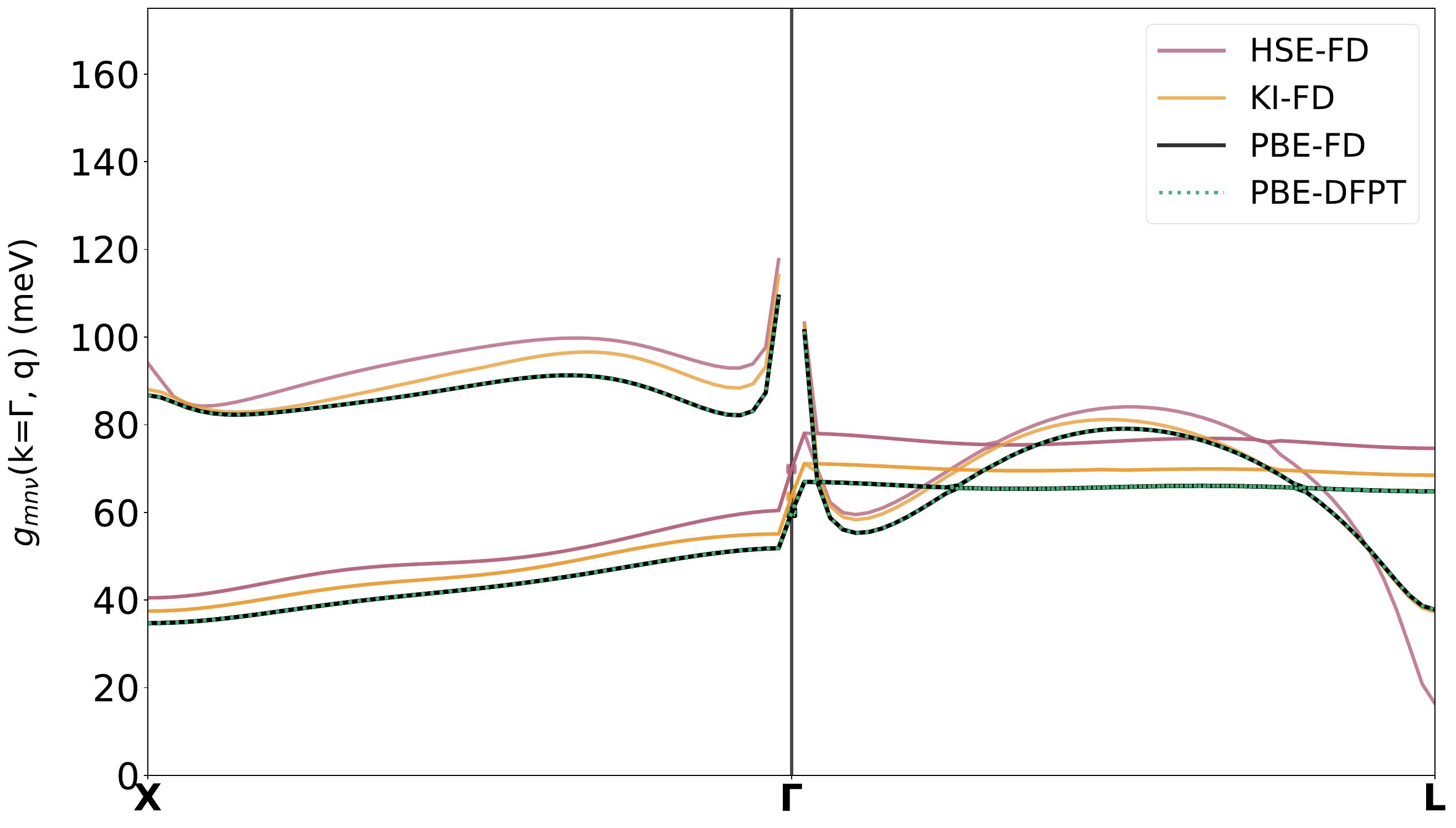}
    \caption{\label{fig:ep_epw_gaas}Electron-phonon matrix elements of GaAs for $n=m=4$ optical branches with different beyond DFT functionals. 
    Interpolation starting from a 4$\times$4$\times$4 supercell.  }
\end{figure}

The phonon band structure of Si with different beyond-DFT methods are shown in Fig.~\ref{fig:ph_epw}. 
The direct calculation of the electron-phonon matrix element for the first 4 unoccupied bands for \textbf{q}=$\boldsymbol{\Gamma}$ and \textbf{k} along the high symmetry line $\boldsymbol{\Gamma}\boldsymbol{L}$ is shown in Fig.~\ref{fig:el_ph_coarse}.
The limited number of points is due to the 4$\times$4$\times$4 supercell used in these direction calculations.
It is worth mentioning that the HSE and KI functionals introduce a smooth enhancement of the coupling in this case, in contrast to $G_0W_0$, which increases and decreases the coupling depending on the band index.
Interpolated on the $\boldsymbol{X}-\boldsymbol{\Gamma}-\boldsymbol{L}$ path using \textsc{EPW}, the electron-phonon coupling of GaAs is presented in Fig.~\ref{fig:ep_epw_gaas}. 
The correction of beyond-DFT methods also behaves like a smooth function.


\newpage
\clearpage

\section*{Average rescreening of the electron-phonon matrix elements}

The main idea behind rescreening lies in the observations that were made on the rate of convergence of mobilities with different functions.
Numerical tests show that the ratio between DFT and beyond-DFT mobility converges more quickly than the mobility itself.
Moreover, the ratio between the DFT and beyond-DFT electron-phonon matrix elements exhibits the same trend. 
Since mobility is inversely proportional to the square of the electron-phonon matrix elements, it is reasonable to assume that the ratio between these quantities is in agreement.
This is illustrated in the table \ref{tab:rescreen}. 
The rescreening factor and the mobility reduction factor agree reasonably well and converge rapidly with supercell size. 

\begin{figure}[ht]
    \centering
    \includegraphics[width=0.7\textwidth]{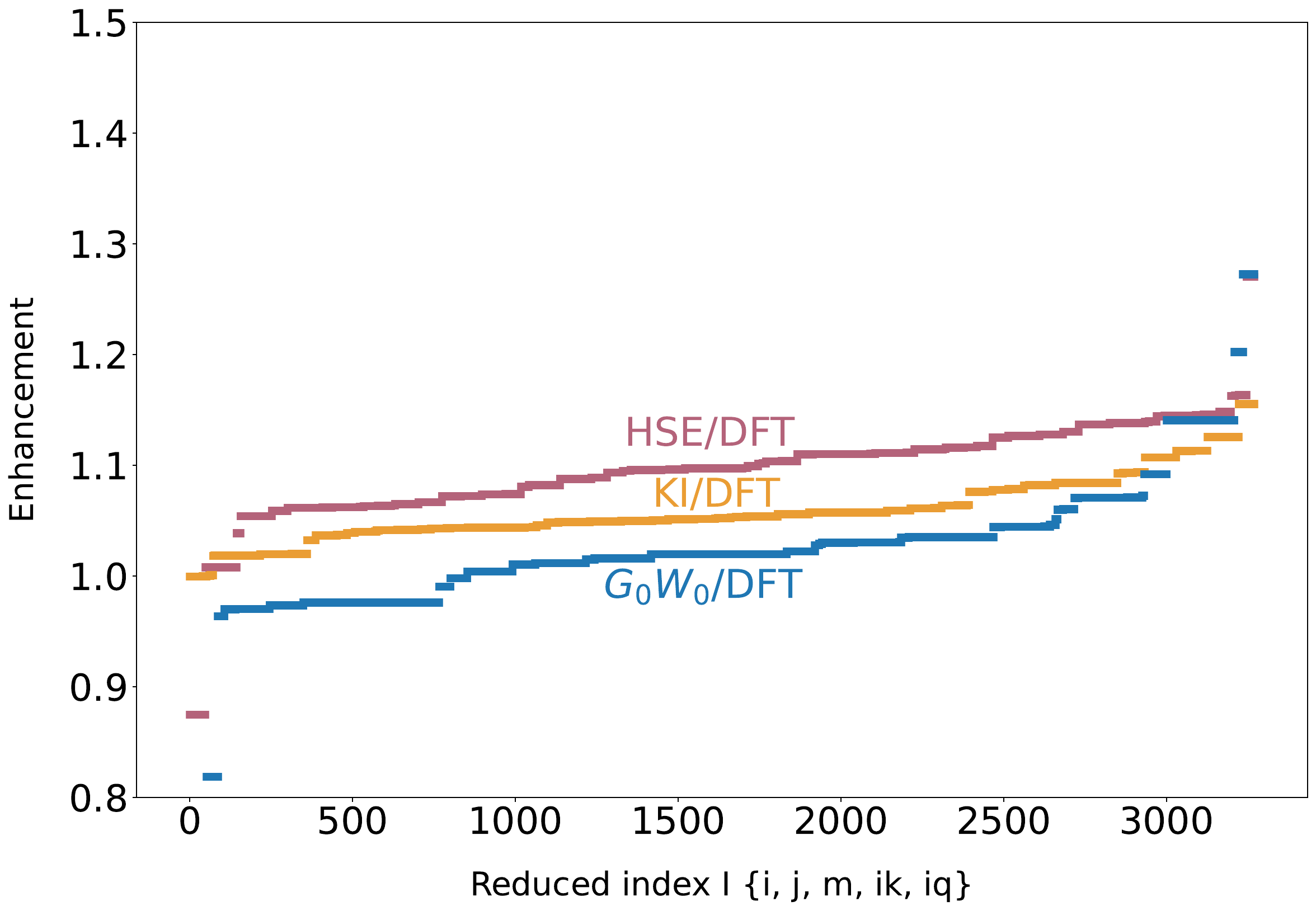}
    \caption{\label{fig:ep_enhancet}  Enhancement of the HSE, KI and $G_0W_0$ electron-phonon matrix elements of Si for the 2$\times$2$\times$2 supercell. 
    The correction of beyond-DFT functionals with respect to DFT is reasonably well approximated by a constant factor for simple systems.}
\end{figure}

\begin{figure}[ht]
    \centering
    \includegraphics[width=0.7\textwidth]{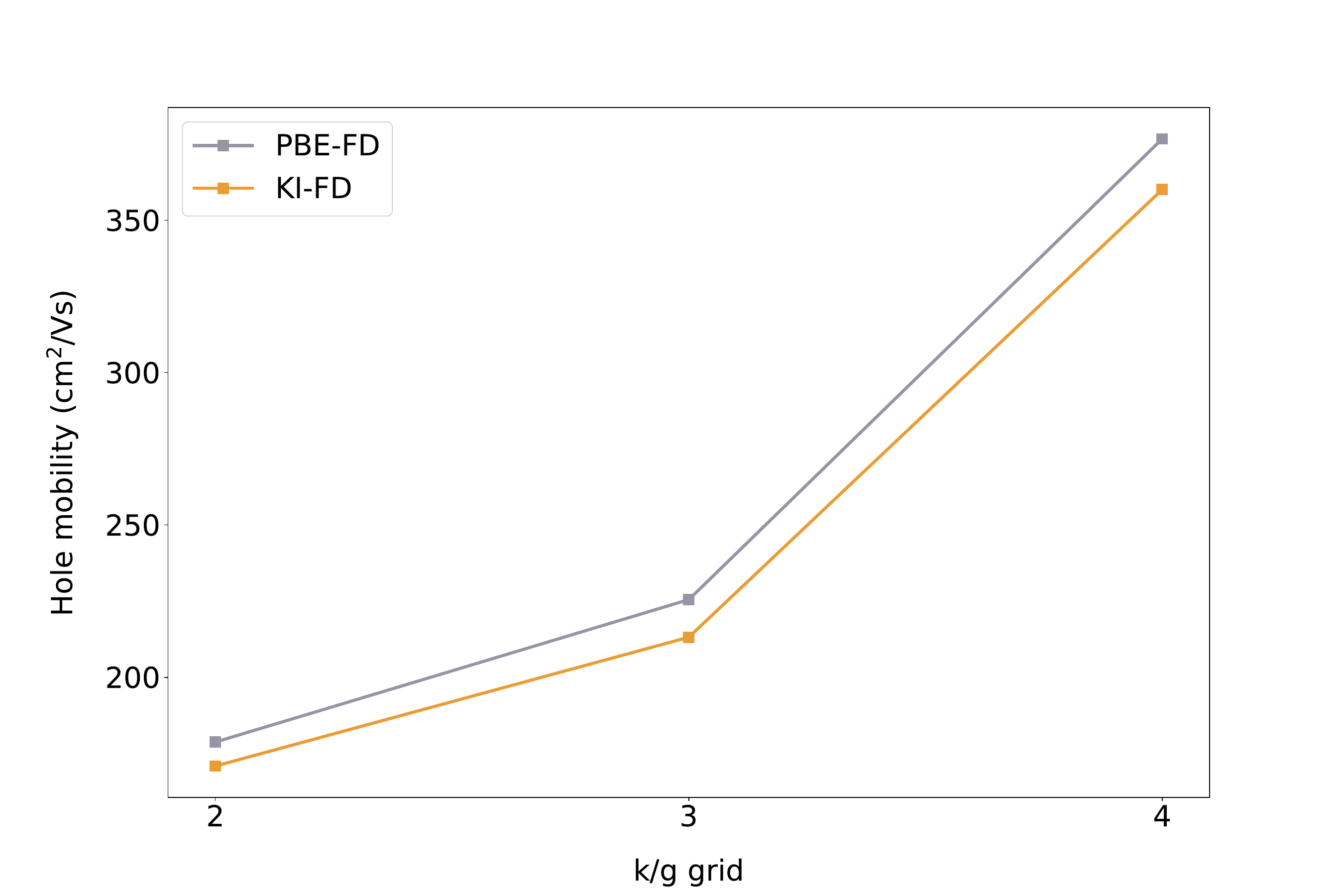}
    \caption{\label{fig:mob_kcw} Convergence of hole mobility of Si with respect to the coarse \textbf{k}/\textbf{q} grid, which is equivalent to the supercell size of the FD approach. 
    Since the effective mass for the holes stays on the DFT level, the only contribution is due to the renormalization of electron-phonon matrix elements.}
\end{figure}

\begin{figure}[ht]
    \centering
    \includegraphics[width=0.7\textwidth]{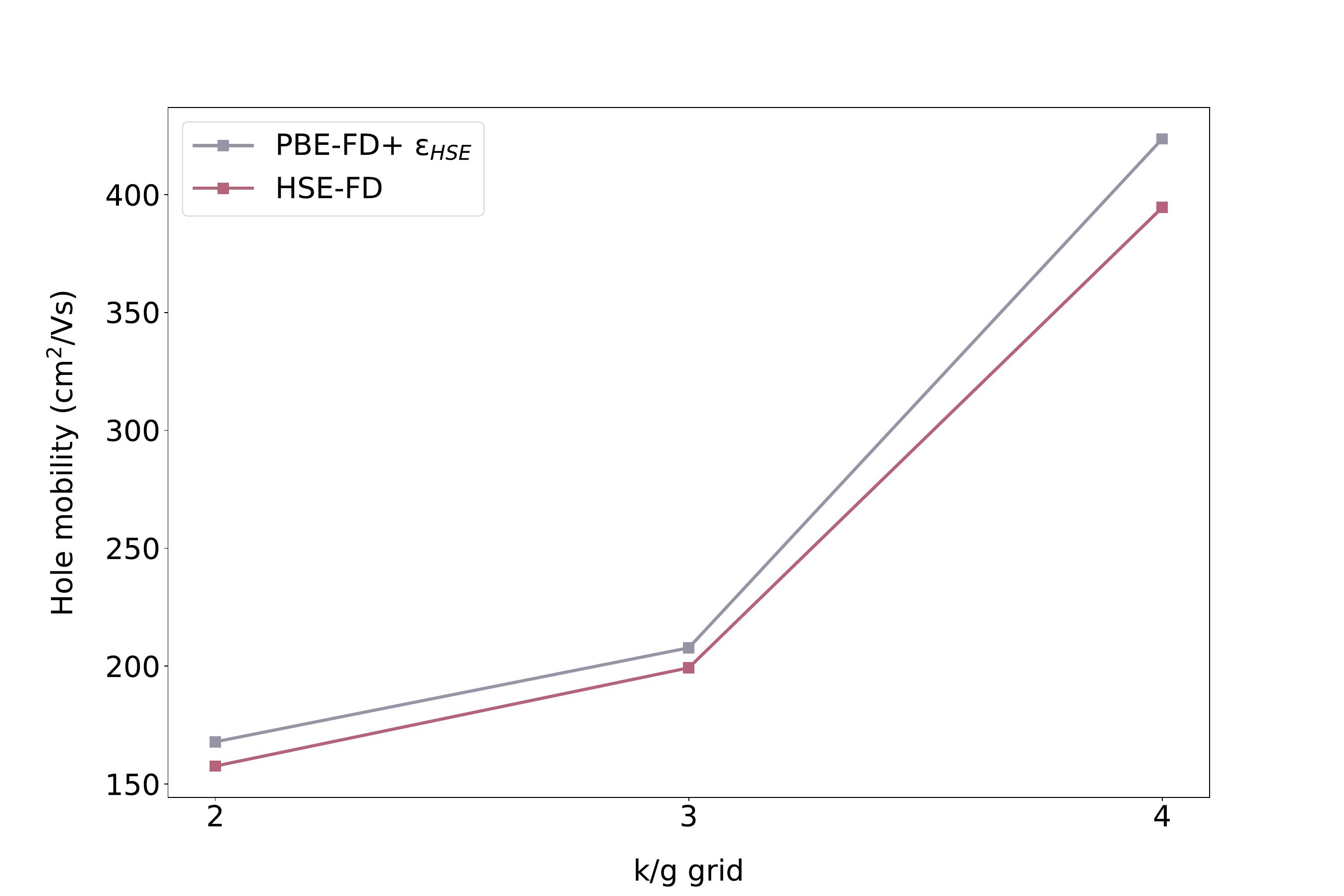}
    \caption{\label{fig:mob_hse} Convergence of hole mobility of Si with respect to the coarse \textbf{k}/\textbf{q} for HSE and DFT with fixed to HSE band structure. 
    The effective mass for the holes is fixed in such a way that the only contribution is due to the renormalization of electron-phonon matrix elements.}
\end{figure}

To elaborate on why this is happening, one could look at the enhancement of the electron-phonon matrix elements due to the beyond-DFT function depicted in Fig.~\ref{fig:ep_enhancet}. 
We see that the rescreening of the electron-phonon matrix elements occurs within a narrow window, which allows for the introduction of a single rescreening factor in the final calculation of the mobility. 
Since the $G_0W_0$ correction requires a sum over a large number of empty states, obtaining converging trends with different supercells is more challenging. 
This is why, for the $G_0W_0$ calculation, we determined rescreening based on the largest supercell, which still remains affordable for the calculation. 
The same trend of rescreening convergence is observed when we look at the mobility of Si, as shown in Figs.~\ref{fig:mob_kcw} and \ref{fig:mob_hse}. for KI and HSE functionals, respectively.
For the occupied bands, the KI correction is a constant shift, and for the comparison between DFT and HSE, we fixed the eigenvalues at the HSE level to neglect the change in the effective mass.
These two figures represent the contribution of only beyond-DFT electron-phonon coupling to mobility. 
As expected, the increased values of the electron-phonon coupling reduce mobility since the latter is inversely proportional to the scattering rates.
The reduction itself remains almost constant, justifying the use of a rescreening factor. 
Since the rescreening factor determined from averaging of electron-phonon couplings could be obtained directly from a coarse grid, we use it instead of the mobility reduction factor in all of the calculations since the latter requires reasonable Wannier interpolation, which is hard to achieve in the case of small supercells. 
Using this rescreening factor allows us to bypass the calculation of electron-phonon matrix elements on prohibitively large supercells, which would otherwise required to match the \textbf{k}-grid necessary to converge the effective masses.

\begin{table}[ht]
    \centering
    \caption{Rescreening of electron-phonon matrix elements and a corresponding reduction of the hole mobility for Si using KI or HSE with respect to DFT(PBE) functional. 
    Since the effective mass stays here on the DFT level, the only contribution is a change in electron-phonon coupling.}
    \label{tab:rescreen}
    \begin{tabular}{|c|c|c|c|c|}
        \hline
        Grid & Mobility reduction (DFT/KI)& $\Bigg|\biggl\langle\dfrac{ g^{\rm KI}_{m n \nu}(\mathbf{k},\mathbf{q})}{ g^{\rm DFT}_{m n \nu}
        (\mathbf{k},\mathbf{q})}\biggl\rangle\Bigg|^2$ & Mobility reduction (DFT/HSE)& $\Bigg|\biggl\langle\dfrac{ g^{\rm HSE}_{m n \nu}(\mathbf{k},\mathbf{q})}{ g^{\rm DFT}_{m n \nu}
        (\mathbf{k},\mathbf{q})}\biggl\rangle\Bigg|^2$  \\
        \hline
        2$^3$ & 1.046 & 1.031 & 1.065 & 1.092\\
        3$^3$ & 1.058& 1.040 & 1.043 & 1.166\\
        4$^3$ & 1.046& 1.038 & 1.074& 1.189\\
        \hline
    \end{tabular}
\end{table}

\newpage
\clearpage
\section*{The effective mass with KI for different variational orbitals.}

\begin{figure}[ht]
    \centering
    \includegraphics[width=0.7\textwidth]{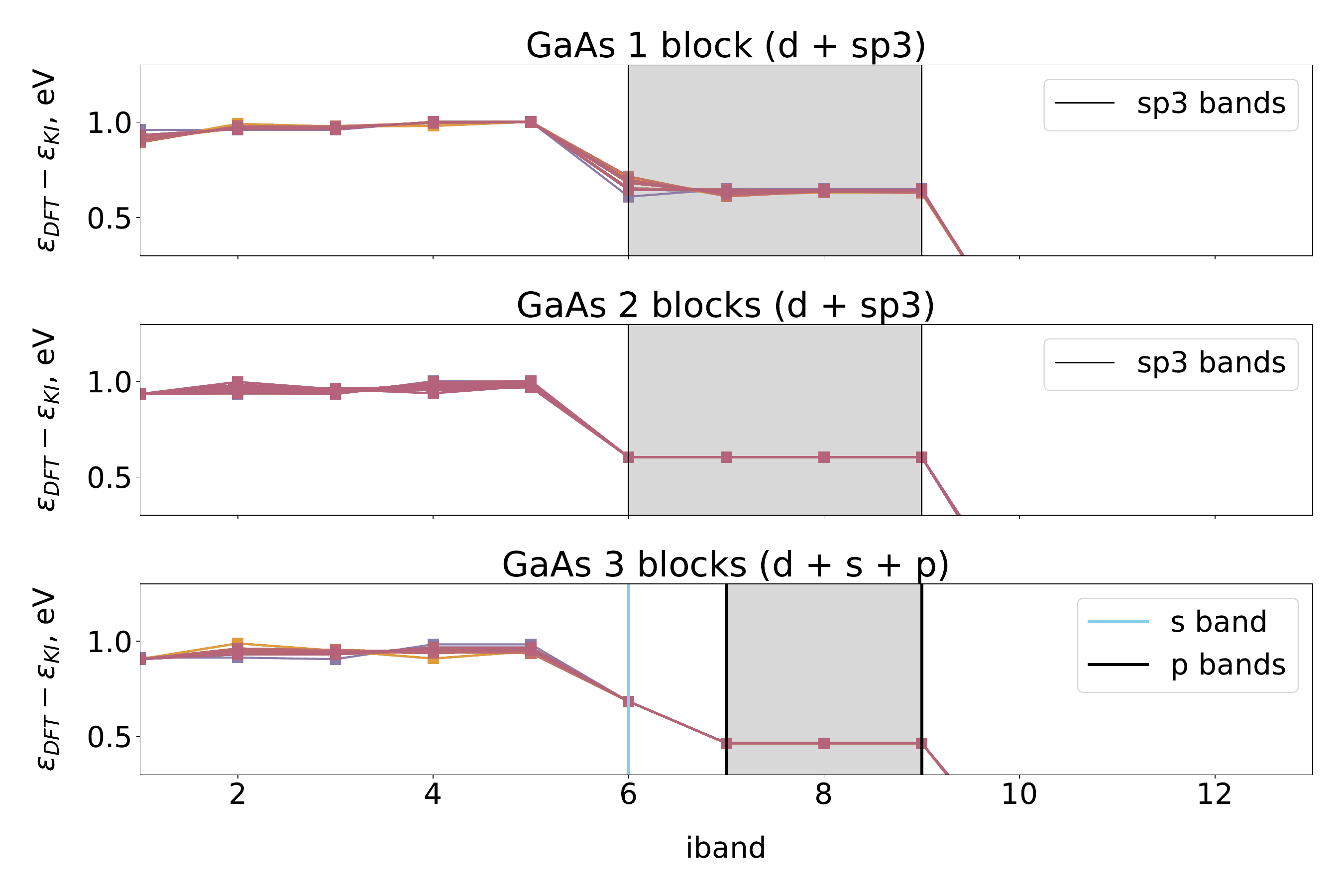}
    \caption{\label{fig:gaas_kcw_eigs} The difference between DFT and Koopmans eigenvalues (shifted so that the VBM matches) for several GaAs states. 
    Different colors represent different \textbf{k}-points. 
    The choice of Wannier functions and the way orbitals are grouped might affect the curvature of the final KI band structure.
    In the calculation where the Wannierization is performed block-by-block and all the orbitals inside the block have the same character (e.g., sp$^{3}$ or p$^{3}$) the correction represents just a constant shift. 
    Note that bands coming from $d$-like Wannier functions will always display a modified dispersion if compared to the DFT one, as in cubic symmetry $d$ states always splits into to set of $t_{2g}$ and $e_g$ states thus having different KI corrections.}
\end{figure}

\begin{figure}[H]
    \centering
    \includegraphics[width=0.7\textwidth]{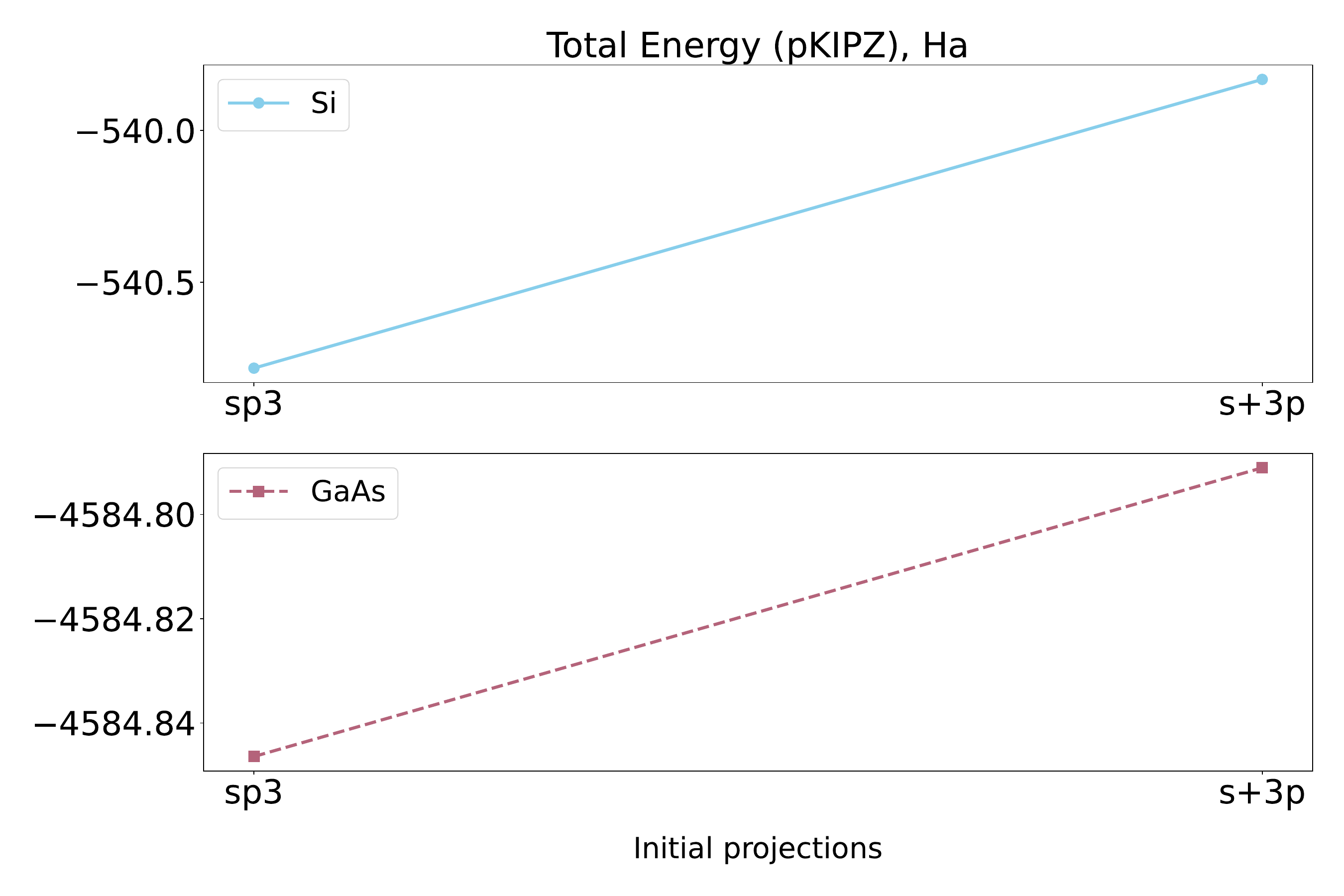}
    \caption{\label{fig:ki_as_kipz} The total energy of pKIPZ functional as a function of initial projection of occupied bands of Si and GaAs. 
    In this case, functional favors to have Wannier functions of the same kind (sp3), delivering minimal energy.}
\end{figure}

We provide here additional details on the structure of the KI Hamiltonian and on how this affects the dispersion ($\mathbf{k}$-dependence) of the KI electronic band structure. 
Assuming the Wannier functions of the KI Koopmans correction comes from a block-by-block Wannierization producing subsets of identical Wannier functions (this is the case in Si where the Wannier functions are 4 $sp^3$-like orbitals), the corrective KI Hamiltonian in this basis can be written as: 
\begin{align}
    \label{eq:KI_ham}
        H^{\rm KI}_{nn'\textbf{k}} =& \bra{\psi^{\rm DFT}_{n\textbf{k}}} \hat{H}^{KI}  \ket{\psi^{\rm DFT}_{n'\textbf{k}}} \\
        =& \varepsilon^{\rm DFT}_{n\textbf{k}} \delta_{nn'} + \sum_{p\textbf{R}} V_p  \bra{\psi^{DFT}_{n\textbf{k}}} \ket{\mathcal{W}_{p\textbf{R}}} \bra{\mathcal{W}_{p\textbf{R}}}\ket{\psi^{\rm DFT}_{n'\textbf{k}}} \\
        =& \varepsilon^{\rm DFT}_{n\textbf{k}} \delta_{nn'}  + \sum_{p_n} V_{p_n} \sum_{m_{p_n}\textbf{R}}   \bra{\psi^{\rm DFT}_{n\textbf{k}}} \ket{\mathcal{W}_{m_{p_n}\textbf{R}}} \bra{\mathcal{W}_{m_{p_n}\textbf{R}}}\ket{\psi^{\rm DFT}_{n'\textbf{k}}} \\
        =& \varepsilon^{\rm DFT}_{n\textbf{k}} \delta_{nn'} + \sum_{p_n} V_{p_{n}} \sum_{m_{p_n}}  U^{\textbf{k}}_{n m_{p_n}} U^{*\textbf{k}}_{ m_{p_n} n'} \\
        =& \varepsilon^{\rm DFT}_{\textbf{k}} \delta_{nn'} + V_{p_n} \delta_{n n'},
\end{align}
where $V_{pn}$ is Koopmans potential, which in the case of occupied bands is simply a constant~\cite{borghi_koopmans-compliant_2014}, but in principle is orbital dependent, $\mathcal{W}_{m_p,\textbf{R}}$ is Wannier function, $m_p$ is the subset of states where $V_{p_n}$ is the same for all orbitals in this subset. 
In this case, the KI corrective term amounts to a block-by-block diagonal correction, meaning that the original DFT states will also be the ones that diagonalize the resulting total KI Hamiltonian and that the final KI eigenvalue will be rigidly shifted by the same amount, $V_{p_n}$, inside each sub-blocks. 
This is exemplified for the case of GaAs in Fig.~\ref{fig:gaas_kcw_eigs}. 
In the general case where orbitals (5 semicore $d$-state and 4 $s$ and $p$ covalence states) are allowed to mix, the simplified correction in Eq.~\eqref{eq:KI_ham} does not apply, and one gets a correction to the band dispersion that is more complex than a constant shift. 
On the other hand, if Wannierization is performed block by block, Eq.~\eqref{eq:KI_ham} applies, and we arrive at a constant shift of the band structure. 
This illustrate the orbital-density-dependence of the Koopmans correction and poses the natural question of which set of Wannier functions we need to choose. 
To answer this question, we recalled the alternative definition of KI as a limit of KIPZ functional with vanishing PZ term. 
Without this definition, any choice of the Wannier function would deliver the same total energy since the KI energy functional for insulating systems is invariant with respect to unitary rotation and would be equally good choice from an energetic point of view (although producing different band structure as discussed above and in Fig.~\ref{fig:gaas_kcw_eigs}).
Turning on even an infinitesimally small PZ term breaks the energy invariance and allow to distinguish between different sets of Wannier functions. 
Figure~\ref{fig:ki_as_kipz} shows the total energy of the pKIPZ functional of Si and GaAs when different sets of Wannier functions are used as proxy for the minimizing orbitals.
Using $sp^3$-like Wannier function as approximated minimizing orbitals produces a lower total KIPZ energy that using a set of 1 $s$-like and 3 $p$-like Wannier functions, and justify the use of the former.  
As a final remark, we state that in the general case when block $m_p$ would have orbitals of different characters, the Koopmans correction to the eigenvalue becomes \textbf{k}-dependent.

%
\putbib[si]
\end{bibunit}

\end{document}